\documentclass{article}

\usepackage{pgfplotstable}
\usepackage{booktabs}
\usepackage{filecontents}
\usepackage{arxiv}
\usepackage{booktabs}
\usepackage[utf8]{inputenc} 
\usepackage[T1]{fontenc}    
\usepackage{hyperref}       
\usepackage{url}            
\usepackage{booktabs}       
\usepackage{amsfonts}       
\usepackage{nicefrac}       
\usepackage{microtype}      
\usepackage{lipsum}		
\usepackage{graphicx}
\usepackage[round]{natbib}
\usepackage{doi}
\usepackage{xcolor}
\usepackage{amsthm,amsmath,amssymb,amsfonts,amssymb,hyperref} 
\usepackage{bm}
\usepackage{cleveref}
\usepackage{float}
\usepackage{multirow}
\usepackage{multibib}
\usepackage{algorithm,algorithmicx,algpseudocode}
\usepackage{blindtext}
\newcommand{\M}{\bm}
\newcommand{\V}{\bm}
\newcommand{\argmax}{\operatornamewithlimits{arg\,max}}
\usepackage{kbordermatrix}

\newcommand{\orcidicon}[1]{\href{https://orcid.org/#1}{\includegraphics[height=\fontcharht\font`\B]{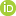}}}

\title{Testing for Asymmetric dependency structures in financial markets: regime-switching and local Gaussian correlation }

\author{ 
        ~\orcidicon{0000-0002-3373-0056} Kristian Gundersen \\
	Department of Mathematics\\
	University of Bergen\\
	Bergen, Norway \\
	\texttt{Kristian.Gundersen@uib.no} \\
	\And	
        ~\orcidicon{0000-0001-6366-5091} Timoth\'ee Bacri \\
        Department of Mathematics\\
	University of Bergen\\
        Bergen, Norway \\
	\texttt{Timothee.Bacri@uib.no} \\
	\And
        ~\orcidicon{0000-0003-3308-0652} Jan Bulla \\
	Department of Mathematics\\
	University of Bergen\\
	Bergen, Norway \\
 	Department of Psychiatry and Psychotherapy\\
	University of Regensburg\\
	Regensburg, Germany\\
	\texttt{Jan.Bulla@uib.no} \\
	\And
        ~\orcidicon{0000-0001-7067-2695}  Sondre Hølleland \\
        Norwegian School of Economics\\
        Bergen, Norway \\
	\texttt{Sondre.Holleland@nhh.no} \\
        \And	
	~\orcidicon{0000-0003-3757-529X} Bård Støve \\
        Department of Mathematics\\
	University of Bergen\\
	Bergen, Norway \\
	\texttt{Bard.Stove@uib.no} 
}

\renewcommand{\shorttitle}{Regime-switching and local Gaussian correlation}

\hypersetup{
pdftitle={Regime shift identification in financial time series},
pdfkeywords={Regime switching, Hidden Markov Models, Local Gaussian Correlation},
}

\begin{document}
\maketitle

\begin{abstract}
This paper examines asymmetric and time-varying dependency structures between financial returns, using a novel approach consisting of a combination of regime-switching models and the local Gaussian correlation (LGC). We propose an LGC-based bootstrap test for whether the dependence structure in financial returns across different regimes is equal. We examine this test in a Monte Carlo study, where it shows good level and power properties. We argue that this approach is more intuitive than competing approaches, typically combining regime-switching models with copula theory. Furthermore, the LGC is a semi-parametric approach, hence avoids any parametric specification of the dependence structure. We illustrate our approach using returns from the US-UK stock markets and the US stock and government bond markets. Using a two-regime model for the US-UK stock returns, the test rejects equality of the dependence structure in the two regimes. Furthermore, we find evidence of lower tail dependence in the regime associated with financial downturns in the LGC structure. For a three-regime model fitted to US stock and bond returns, the test rejects equality of the dependence structures between all regime pairs. Furthermore, we find that the LGC has a primarily positive relationship in the time period 1980-2000, mostly a negative relationship from 2000 and onwards. In addition, the regime associated with bear markets indicates less, but asymmetric dependence, clearly documenting the loss of diversification benefits in times of crisis.
\end{abstract}


\section{Introduction}

Dependence between asset returns is important in many aspects in finance, in particular for portfolio theory, where the aim is to allocate assets by maximizing the expected return of the portfolio while minimizing its risk, for instance measured by the standard deviation. The rule is simple: weakly correlated assets are good for diversification, but highly correlated assets should be avoided. The crucial assumption is that the asset returns follow a joint-Gaussian distribution in this classical mean-variance approach, see \citet{mark:1952}. The advantage of the Gaussian approach for modelling asset returns is that it is straightforward. Solely based on means and covariances, it leads to a complete theoretical framework in the considered multivariate framework. 

However, the restrictive nature of the Gaussian distribution approach is well-documented, as asymmetries are often found in the distribution of financial returns,  \citep[see, for example,][]{silv:gran:2001, longin:2001, Ang2002b, Hong2007, Okimoto2008, Chollete2009, aas:czad:frig:bakk:2009, stov:tjos:2014, bernardi2017multiple, bens:2018}. One main finding opposing the Gaussian assumption is the often stronger dependence between returns of financial assets during periods of market downturn or crashes (often called «bear markets»), and less dependence in stable or increasing markets (often called «bull markets»), hence time-varying dependency structures are observed. Another well-known asymmetry is the skewness in the distribution of individual asset returns. This has led to the conclusion that the Gaussian distribution is not well-founded empirically \citep[see, e.g.,][]{rydb:shep:2000}.

There are several methods for studying asymmetry of financial returns. \citet{silv:gran:2001} looked at various quantile estimation methods, and \citet{longin:2001} employed extreme value theory to show that there is a bear market effect, but no bull effect, for monthly data. \citet{Okimoto2008}, \citet{Rodriguez2007} and \citet{bens:2018} have employed regime-switching copulas to study asymmetric dependence for various international stock indices. Moreover, for instance \citet{aas:czad:frig:bakk:2009} and \citet{Niko2012} have used vine copulas (also called the pair-copula construction) to model multivariate financial return data. Related works are \citet{Ang2002a} and \citet{Ang2002b}, who have based themselves on Markov regime structures with ARCH/GARCH modeling. Selected further references to the modelling of financial returns using regime-switching models are \citet{hardy2001}, \citet{bulla2006}, and \citet{maruotti2019}.   

Recently, factor copulas have been introduced for modeling dependence in high dimensions, see e.g. \citet{patton2017}. Also, \citet{christ:2012} model the correlation among a large set of countries with a dynamic asymmetric copula (or DAC), concluding that correlations have increased markedly in both developed markets and emerging markets over the past decades. Another way of modeling time-varying correlation, is the use of the very popular dynamic conditional correlation (DCC) estimators, which possesses the flexibility of univariate GARCH models without the complexity of conventional multivariate GARCH, see \citet{engle:2002}.

A common feature for many of the alternative approaches mentioned above is that one ends up with one or more parameters that have a rather indirect interpretation as a measure of dependence. In this respect, correlation has a more natural basis. Local Gaussian correlation  \citep[LGC, see][]{tjostheim2013local} is a local dependence measure capable of revealing asymmetric dependence, and interpretable as the standard correlation. It has been successfully applied to analyze dependence structures between asset returns \citep[see, e.g.,][]{stov:tjos:2014, stov:tjos:huft:2014, bampinas:2017, nguyen:2020}. However, none of these studies have examined the time-varying local Gaussian correlation in a structured way, and the aim of this paper is to close this gap. Hence, in this paper we combine the use of local Gaussian correlation with regime-switching models, and propose a formal test for equality of   dependence structures in financial markets across regimes, taking into account the existence of any asymmetric dependence structures. We will not limit ourselves to testing across only two regimes, even though many studies \citep[e.g.][]{Ang2002a, Okimoto2008} document that there are typically two distinct regimes observed in financial returns series. The test procedure is related to the test for financial contagion presented in \cite{stov:tjos:huft:2014}. However, the test developed in this paper is a more general test for examining whether dependency structures between financial returns are different across regimes, and not only focusing on a "stable" time period and a "crisis" time period. Furthermore, the proposed test in this paper bases on the whole LGC map, and is not limited to testing on the diagonal elements. The advantages of this extension will become clear in the empirical analysis of this paper. 

The organisation of the paper is as follows. Section 2 briefly reviews LGC and regime-switching models. Section 3 presents our methodological set-up, including a nonparametric bootstrap test for asymmetric dependence across regimes, and examining its level and power properties in a Monte Carlo study. In Section 4, we illustrate the approach by performing several empirical analyses by the example of different financial return data sets, while Section 5 offers
some conclusions.

\section{Methodology}

In this section we briefly review the main theory of the local Gaussian correlation (LGC) and regime-switching models. Book length treatments are found in \cite{book} and \cite{zucchini}, respectively.

\subsection{Local Gaussian correlation}\label{LGC}

This paper relies on the relatively recently developed dependence measure LGC, introduced by \cite{tjostheim2013local}. This is a local characterization of dependence, and the underlying idea has also been extended to several different situations. These include a test of independence \citep{bere:tjos:2014, laca:tjos:2017, laca:tjos:2018}, density and conditional density estimation \citep{otne:tjos:2017,otne:tjos:2018}, a local Gaussian partial correlation \citep{otneim2021}, local Gaussian spectral \citep{jordanger2022nonlinear} and cross-spectrum estimation \citep{jordanger2023local}. Finally, the relationship between the local Gaussian correlation and different copulas has been studied in \cite{berentsen2014recognizing}. A thorough overview of the local Gaussian approximation approach can be found in \cite{book}. For completeness, we present the local Gaussian correlation in a standard way, and we note that this section closely follows the presentation of the LGC in \cite{tjos:otne:stov:2021}. 

Finally, as already mentioned in the introduction, the local Gaussian correlation has been used in several studies examining the dependence structure between asset returns, testing for financial contagion, and utilzed in portfolio allocation, see e.g. \cite{stov:tjos:2014}, \cite{stov:tjos:huft:2014}, \cite{bampinas:2017}, \cite{nguyen:2020}, \cite{sleire:2021} and \cite{ming:2022}, but not in conjunction with regime-switching models, which is the focus of this paper.

\subsubsection{Definition}

Let $\bm R=(R_1,R_2)\in \mathbb{R}^2$ represent the stochastic return variable of two risky assets with bivariate density $f$ and let $\bm r=(r_1,r_2)\in\mathbb R^2$ denote a realisation of said variable. For simplicity we drop the time index here. We approximate $f$ locally in each point $\bm x=(x,y)\in\mathbb R^2$ by a Gaussian bivariate density, \(\psi_{\bm x}(\V v)\), where \( \V v = (v_1,v_2)\) are running variables. Let \(\bm \mu(\bm x) = (\mu_1(\bm x),\mu_2(\bm x))\) be the mean vector in the normal distribution having density \(\psi_{\bm x}\), \(\bm\sigma(\bm x) = (\sigma_1(\bm x),\sigma_2(\bm x))\) is the vector of standard deviations, and \(\rho(\bm x)\) is the correlation coefficient in the normal distribution \(\psi_{\bm x}\). The approximating density is then given as
\begin{align}
&\psi_{\bm x} = \psi(\bm v,\mu_1(\bm x),\mu_2(\bm x),\sigma_1^2(\bm x),\sigma_2^2(\bm x),\rho(\bm x)) = \frac{1}{2\pi\sigma_1(\bm x)\sigma_2(\bm x)\sqrt{1-\rho^2(\bm x)}} \nonumber \\
& \qquad\times \exp\Big[-\frac{1}{2} \frac{1}{1-\rho^2(\bm x)}\Big(\frac{(v_1-\mu_1(\bm x))^2}{\sigma_1^2(\bm x)}-2\rho(\bm x)\frac{(v_1-\mu_1(\bm x))(v_2-\mu_2(\bm x))}{\sigma_1(\bm x)\sigma_2(\bm x)} \nonumber \\ & \qquad\qquad +\frac{(v_2-\mu_2(\bm x))^2}{\sigma_2^2(\bm x)}\Big) \Big]. 
\label{eq:bivariate-normal}
\end{align}
Moving to another point \(\bm x'\) results in another approximating normal distribution \(\psi_{\bm x'}\) which depends on a new set of parameters \((\mu_1(\bm x'),\mu_2(\bm x'), \sigma_1(\bm x'), \sigma_2(\bm x'), \rho(\bm x'))\). One exception to this is the case where \(f\) itself is Gaussian with parameters \((\mu_1,\mu_2, \sigma_1, \sigma_2 ,\rho)\), in which case \((\mu_1(\bm x),\mu_2(\bm x), \sigma_1(\bm x), \sigma_2(\bm x), \rho(\bm x)) \equiv (\mu_1, \mu_2, \sigma_1, \sigma_2, \rho)\).

The population parameter vector \(\bm \theta(\bm x) \stackrel{\textrm{def}}{=} (\mu_1(\bm x),\mu_2(\bm x),\sigma_1(\bm x),\sigma_2(\bm x), \rho(\bm x))\) is obtained by minimizing the local penalty function 
 measuring the difference between \(f\) and \(\psi_{\bm x}\). It is defined by
\begin{equation}
q = \int K_{\V b}(\V v-\bm x)[\psi(\V v,\V \theta(\bm x))-\ln \{\psi (\V v,\V \theta(\bm x))\}f(\V v)] \textrm{d} \V v
\label{eq:Kullback-distance}
\end{equation}
where \(K_{\V b}(\V v-\bm x) = (b_1b_2)^{-1}K_1(b_1^{-1}(v_1-x))K_2(b_2^{-1}(v_2-y))\) is a product kernel with bandwidths $\V b = (b_1, b_2)$. As is seen in Hjort and Jones (1996, pp 1623-1624), the expression in \eqref{eq:Kullback-distance} can be interpreted as a locally weighted Kullback-Leibler distance from \(f\) to \(\psi(\cdot,\V \theta(\bm x))\). Hence, the minimizer \(\V \theta_{\V b}(\bm x)\) (which also depends on \(K\)) should be a solution of
\begin{equation}
\int K_{\V b}(\V v-\bm x)\frac{\partial}{\partial \theta_j} [\ln\{\psi(\V v,\V \theta(\bm x))\}f(\V v)-\psi(\V v,\V \theta(\bm x))] \textrm{d} \V v = 0, \;\;j=1,\ldots,5.
\label{eq:score-equation}
\end{equation}\\
In the first step, we define the population value \(\V \theta_{\V b}(\bm x)\) as the minimizer of \eqref{eq:Kullback-distance}, assuming that there is a unique solution to \eqref{eq:score-equation}. The definition of \(\V \theta_{\V b}(\bm x)\) and the assumption of uniqueness are essentially identical to those used in \cite{hjor:jone:1996} for more general parametric families of densities.

In the next step, we let \(\V b \to \V 0\) and consider the limiting value \(\V \theta(\bm x) = \lim_{\V b\rightarrow \V 0} \V \theta_{\V b}(\bm x)\). This is in fact considered indirectly by \cite{hjor:jone:1996} and more directly in \cite{tjostheim2013local}, both using Taylor expansion arguments. In the following we assume that there exists a limiting value \(\V \theta(\bm x)\) independent of \(\V b\) and \(K\). 

\subsubsection{Estimation and likelihood function}

When estimating \(\bm \theta(\bm x)\) and \(\bm \theta_{\bm b}(\bm x)\) we have to use a neighborhood with a finite bandwidth, which is in analogy to nonparametric density estimation. The estimate \(\widehat{\bm \theta}(\bm x) = \widehat{\bm \theta}_{\bm b}(\bm x)\) is then obtained from maximizing a local likelihood. Given observations \(\bm R_1,\ldots,\bm R_T\), the local log likelihood is determined by
\begin{align}
L(\bm R_1,\ldots,\bm R_T,\bm \theta(\bm x)) &= T^{-1}\sum_i K_{\bm b}(\bm R_i - \bm x)\log \psi(\bm R_i,\bm \theta(\bm x)) \nonumber \\ & \qquad\qquad\qquad - \int K_b(\bm v - \bm x)\psi(\bm v, \bm \theta(\bm x))\textrm{d} \bm v.
\label{eq:LGC-35}
\end{align}
When \(\bm b \to \infty\), the last term has 1 as its limiting value, and the likelihood reduces to the ordinary global likelihood. This last term is essential, as it implies that \(\psi(\bm x, \bm \theta_{\bm b}(\bm x))\) is not allowed to stray far away from \(f(\bm x)\) as \(\bm b \to \bm 0\). Indeed, with the notation
\begin{equation}
u_j(\cdot,\bm \theta) \stackrel{\textrm{def}}{=} \frac{\partial}{\partial \theta_j} \log \psi(\cdot,\bm \theta),
\label{eq:LGC-defu}
\end{equation}
and assuming ${\rm{E}} (K_{\bm b} (\bm R_i - \bm x) \log \psi(\bm R_i, \bm \theta_{\bm b}(\bm x))) < \infty$, we have almost surely

\begin{align}
\frac{\partial L}{\partial \theta_j} &= T^{-1}\sum_i K_{\bm b}(\bm R_i - \bm x)u_j(\bm R_i, \bm \theta_{\bm b}(\bm x)) \nonumber \\ 
& \qquad - \int K_{\bm b}(\bm v - \bm x)u_j(\bm v,\bm \theta_{\bm b}(\bm x))\psi(\bm v,\bm \theta_{\bm b}(\bm x)) \textrm{d} \bm v \nonumber \\
&\to \int K_{\bm b}(\bm v - \bm x)u_j(\bm v,\bm \theta_{\bm b}(\bm x))[f(\bm v) - \psi(\bm v, \bm \theta_{\bm b}(\bm x))] \textrm{d} \bm v.
\label{eq:LGC-4}
\end{align}
by the law of large numbers, or by the ergodic theorem in the time series case. Setting the expression in the first line of \eqref{eq:LGC-4} equal to zero yields the local maximum likelihood estimate \(\widehat{\bm \theta}_{\bm b}(\bm x)\) (\(=\widehat{\V\theta}(\bm x)\)) of the population value \(\bm \theta_{\bm b}(\bm x)\) (and \(\bm \theta(\bm x)\) which satisfies \eqref{eq:score-equation}). Hence, for each point \(\bm x\), also referred to as gridpoints in the sequel, we obtain an estimate for the correlation in that point, $\hat{\rho}(\bm x)$, which we call the local Gaussian correlation. Maximizing the likelihood in several gridpoints, thus results in several estimates of the local Gaussian correlations, that constitutes what we call an LGC map in the sequel. Hence, we are thus able to describe any potential asymmetric dependence patterns by this map of locally estimated correlations.

An asymptotic theory has been developed in \cite{tjostheim2013local} for \(\widehat{\V \theta}_{\V b}(\bm x)\) for the case that \(\V b\) is fixed and for \(\widehat{\V \theta}(\bm x)\) in the case that \(\V b \to \V 0\). The first case is much easier to treat than the second one. In fact for the first case the theory of \cite{hjor:jone:1996} can be used almost directly, although it is extended to the ergodic time series case in \cite{tjostheim2013local}. In the case that \(\V b\rightarrow \V 0\), this leads to a slow convergence rate of \((n(b_1b_2)^{3})^{-1/2}\), which is the same convergence rate as for the estimated dependence function treated in \cite{jone:1996}.

As already mentioned, the local estimates depend on the smoothing device - the bandwidth vector $\V b$ and a specific choice of the kernel function, $K$. There are various ways of selecting the bandwith parameter $\V b$,  \citep[see, e.g.][]{otne:tjos:2018, bere:tjos:2014, stov:tjos:huft:2014}. 

\subsubsection{Multivariate case}

We have thus far concentrated on the bivariate case, in which we estimate a single local Gaussian correlation map based on a bivariate sample, and in the present paper we restrict ourselves to this situation. However, it is in principle straightforward to extend to the case of more than two variables. Assume that we observe a multivariate sample $\V R_t = \{R_{1t}, \ldots, R_{pt}\}$, $t=1,\ldots, T$ with dimension $p>2$. We can then estimate the $p\times p$ local correlation matrix $\M \rho(\bm x) = \{\rho_{k\ell}(\bm x)\}$, $1\leq k< \ell \leq p$, as well as the local means and local variances $\V \mu(\bm x) = \{\mu_1(\bm x), \ldots, \mu_p(\bm x)\}$ and $\V \sigma(\bm x) = \{\sigma_1(\bm x), \ldots, \sigma_p(\bm x)\}$ by maximizing the local likelihood function \eqref{eq:LGC-35}. The precision of such estimates, however, deteriorates quickly as the dimension $p$ grows, due to the curse of dimensionality. 

But, a simplifying technique that reduces the complexity of this estimation problem, introduced by \cite{otne:tjos:2017}, is to estimate each local correlation $\rho_{k\ell}(\V z)$ as a bivariate problem by only considering the corresponding pair of observation vectors $\{R_{kt}, R_{\ell t}\}$, $t=1,\ldots,T$. Thus, we reduce the $p$-variate problems of estimating the local parameters depending on all coordinates, to a series of bivariate problems of estimating pairwise local correlations depending on their respective pairs of coordinates. In this way, we obtain a simplification that is analogous to an additive approximation in nonparametric regression. For more details regarding this pairwise modeling approach, see \cite{otne:tjos:2017}. 

\subsection{Regime-switching models - hidden Markov models}
\label{sec:HMMs}

In this paper, we employ a regime-switching model - also known by the name hidden Markov model (HMM) - to allow for switching between different regimes (or states, used interchangeably). First used in speech recognition \citep[see, e.g.,][]{baum, fredkin, gales}, these models are now employed in ecology \citep{mcclintock}, biology and bioinformatics \citep{schadt, durbin, eddy}, finance \citep{hamilton, quandta, anga}, and many other fields.

The two commonly used estimation procedures for HMMs are Direct Numerical Maximization (DNM) of the likelihood as introduced by \citet{turner} and later detailed by \citet{macdonald}, and Expectation Maximization (EM)-type algorithms as introduced by \citet{bauma, dempster}. Each procedure possesses advantages and downsides, for example a main difference is the robustness of the EM algorithm towards poor initial values. More details and a comparison of both approaches are discussed in \citet{bulla}, who also describe a hybrid approach combining both algorithms. For simplicity, we choose to adopt the DNM approach, as it is easier to adapt to different situations. In addition, we employ the Template Model Builder \citep[TMB, ][]{kristensen2015tmb} package in R to accelerate the estimation process. We refer to \citep{bacri2022gentle, bacri2023computational}  for a tutorial on TMB with HMMs using DNM, along with detailed Poisson, Gaussian, and multivariate Gaussian examples and an overview of suitable optimization algorithms.

\hypertarget{definition}{%
\subsubsection{Definition}
\label{definition}}

Roughly speaking, HMMs are characterized by switching between \(C\) so-called conditional distributions (or regimes) in time, where the switching process is governed by latent Markov chain. Similarly to the notation from \Cref{LGC}, we let \(\{\bm{R_t}: t = 1, \ldots, T\}\) and \(\{S_t : t = 1, \ldots, T\}\) denote respectively an observed multivariate time series and the states of a hidden (unobserved) Markov chain, where \(t\) denotes the (time) index ranging from one to \(T\). For the purposes of this paper, the hidden Markov chain is assumed homogeneous, irreducible and aperiodic.

We define our \(C\)-state Gaussian HMM through bivariate Gaussian conditional distributions, i.e., the probability density function equals

\begin{equation*}
p_i(\bm{r}) = \text{P}(\bm{R}_t = \bm{r} \vert S_t = i) = \frac{1}{\sqrt{2 \pi \det (\bm{\Sigma}_i)}} \exp\left(-\frac{1}{2} (\bm{r} - \bm{\mu}_i)' \bm{\Sigma}_i^{-1} (\bm{r} - \bm{\mu_i}) \right),
\end{equation*}

with parameters \((\bm{\mu}_i, \bm{\Sigma}_i)\), where \(i = 1, \ldots, C\). Any conditional distribution could be used, but as we are mainly interested in the difference in dependence structures across regimes, we use the Gaussian distribution for convenience.
The latent Markov chain of the HMM is characterized by a transition probability matrix (TPM) that we denote \(\bm{\Gamma} = \{\gamma_{ij}\}\). We assume ergodicity of the chain, which implies existence and uniqueness of the stationary distribution as the limiting distribution, which we denote \(\bm{\delta}\). For more details on these results, we refer to \citet[Lemma 6.3.5 on p.~225 and Theorem 6.4.3 on p.~227]{grimmett} and \citet[p.~394]{feller}.

\hypertarget{Likelihood-function}{%
\subsubsection{Likelihood function}
\label{likelihood-function}}

Estimation of the HMM via DNM requires computation of the likelihood. Let \(\bm{R}^{(t)} = \{\bm{R}_1, \ldots, \bm{R}_t \}\) and \(\bm{r}^{(t)} = \{\bm{r}_1, \ldots, \bm{r}_t \}\) denote the 'history' of the observed process \(\bm{R}_t\) and of the observations \(\bm{r}_t\), respectively, with \(t\) denoting the time ranging from one to \(T\). Moreover, \(\bm{\zeta}\) denotes the vector of model parameters. With this notation, the likelihood of the observations can be written as
\begin{equation}
\label{eq:hmm_likelihood}
L(\bm{\zeta}) = \text{P}(\bm{R}^{(T)} = \bm{r}^{(T)}) = \bm{\delta} \mathbf{P}(\bm{r}_1) \bm{\Gamma} \mathbf{P}(\bm{r}_2) \bm{\Gamma} \mathbf{P}(\bm{r}_3) \ldots \bm{\Gamma} \mathbf{P}(\bm{r}_T) \bm{1}',
\end{equation}
where the \(C\) conditional probability density functions evaluated at \(\bm{r}\) can be represented as the diagonal matrix
\begin{equation*}
\mathbf{P}(\bm{r}) = \begin{pmatrix}
p_1(\bm{r})    &         &         & 0\\
          & p_2(\bm{r})  &         &\\
          &         & \ddots  &\\
0         &         &         & p_C(\bm{r})
\end{pmatrix},
\end{equation*} and \(\bm{1}\) denotes a vector of ones and \(\bm{\delta}\) denotes the stationary distribution. When \(\bm{r}\) is a missing observation, one can set \(p_i(\bm{r}) = 1\) \(\forall i\), hence \(\mathbf{P}(\bm{r})\) becomes the unity matrix as explained by
\citet[p.~40]{zucchini}. We choose to set the first term of the likelihood - the so-called initial distribution - to \(\bm{\delta}\). Note, however, that it is also possible to freely estimate the initial distribution \citep[Section 2.3.2 Proposition 1 p.~37]{zucchini}.

\subsubsection{State inference}
\label{state-inference}

Once an HMM has been estimated, it is possible to determine the sequence of most likely states of the data set. These states can be inferred by a method known as local decoding through so-called smoothing probabilities, as detailed in \citet[Chapter 5]{zucchini}. Let us define the so-called forward
\[
\alpha_t(i) = \text{P}(\V R^{(T)} = \V r^{(T)} \vert S_t = i)
\] and backward probabilities
\[
\beta_t(i) = \text{P}(R_{t+1} = r_{t+1}, R_{t+2} = r_{t+2}, \ldots, R_{T} = r_{T} \vert S_t = i).
\]
Then, the smoothing probabilities $\text{P}(S_t = i \vert \bm{R}^{(T)} = \bm{r}^{(T)})$ equal
\[
\text{P}(S_t = i \vert \bm{R}^{(T)} = \bm{r}^{(T)}) = \frac{\alpha_t(i) \beta_t(i)}{L(\bm{\psi})}
\]
for \(i = 1, \ldots, C\) and \(t = 1, \ldots, T\),
and correspond to the conditional probability of being in state \(i\) at time \(t\) given all observations.
The most probable state \(i_t^*\) at time $t$ then directly follows from  the maximal smoothing probability over all possible states through 
\[
i_t^* = \argmax_{i \in \{1, \ldots, m \}} \text{P}(S_t = i \vert \bm{R}^{(T)} = \bm{r}^{(T)}).
\]

\section{Comparing dependence across regimes}\label{sec:test}

The main purpose of the suggested approach is to combine a regime-switching model with the local Gaussian correlation to describe regimes in financial returns. Furthermore, to test whether the dependence structure between the returns differs across the regimes. The method is a step-wise procedure, which we explain in the following.
 
We consider a bivariate sample $\V R_t = \{R_{1t}, R_{2t} \}$, where $t=1,\ldots, T$. Each bivariate observation at time $t$ is classified into a regime $c_t \in \{ 1, \ldots ,C \}$ with a fitted HMM to the whole sample. The classified bivariate observation at time $t$ is thus denoted $\V R_t^{c_t} = \{R_{1t}^{_t}, R_{2t}^{c_t} \}$, where $c_t$ denotes the regime of the specific observation at $t$. Observations with equal regimes constitute a subset of $\V R$, i.e. $\V R^{c_t} \subset \V R$. Our goal is to examine the dependency structure over the different regimes $c_t$ in $\V R^{c_t}$ with the LGC measure, as decribed in \Cref{LGC}. Note that one usually needs to perform a filtration of the data to remove dependence over time, and to remove volatility effects, this is further elaborated in Section \ref{GARCH_filtering}.

For all observations within a specific regime $c$, we can estimate the $2 \times 2$ local correlation matrices $\M \rho_{c}(\V x)$ in the grid point $(x,y)$ by maximizing the local likelihood function \eqref{eq:LGC-35}. 
   
Furthermore, using several gridpoints, we estimate the LGC map for each of the $C$ different regimes. We can therefore proceed to examine whether the dependency structure for the different regimes of the time series are equal or not. Hence, we propose a bootstrap test procedure, and we note that similar test procedures are often used in a nonparametric setting, e.g. for testing difference between quantities in nonparametric regressions, see e.g. \citet{hall1990bootstrap} or \citet{vilar2007bootstrap}. 

\subsection{Bootstrap test}

To accommodate for any asymmetric dependence structures, we test on the entire LGC map. We use $i,j$ as notation for specifying the gridpoints such that $\M x_{ij} = (x_i, y_j)$, where $i=1, \ldots ,n$ and $j = 1 \ldots ,n$. The test we propose here is similar to \citet{stov:tjos:huft:2014}, who developed a bootstrap test for contagion between financial time series. Where \citet{stov:tjos:huft:2014} considered the diagonal elements, $(x_i, y_i)$, this bootstrap procedure considers the entire grid $(x_i, y_j)$. Performing the test on the entire grid instead of on the diagonal elements ensures robustness to reveal non-linear dependencies between the LGC maps of the different regimes. The test on the entire grid $\M x_{ij}$ and arbitrary many regimes $C$ can be formulated with the following null and alternative hypothesis
\begin{align*}
 H_0: \quad & \M \rho_{1}(x_i, y_j) = \M \rho_{2}(x_i, y_j) = \ldots = \M \rho_{C}(x_i, y_j) \quad \text{for} \quad i,j=1,\cdots,n (\text{no difference in dependence across regimes})\\
 H_1:\quad & \M \rho_{1}(x_i, y_j) \neq \M \rho_{2}(x_i, y_j) \neq \ldots \neq \M \rho_{C}(x_i, y_j) \quad \text{for} \quad i,j=1,\cdots,n (\text{difference in dependence across regimes})
\end{align*}
The bootstrap method works as follows. From the classified observations $\{\V R^{c_1}_1, \V R^{c_2}_2 \ldots \V R^{c_T}_T \}$, we draw randomly and with replacement a re-sample $\{\V R^{c_1*}_1, \V R^{c_2*}_2 \ldots \V R^{c_T*}_T \}$ and by gathering observations classified to the same regime $c \in \{ 1, \ldots ,C \}$, compute $\hat{\M \rho}_{1}^*(x_i, y_j), \hat{\M \rho}_{2}^*(x_i, y_j), \ldots, \hat{\M \rho}_{C}^*(x_i, y_j)$ on the grid $\M x_{ij}$ for $i,j=1, \ldots ,n$. With $C$ classes we can pairwise test between different regimes. Excluding to test between equal regimes and any perturbations there are $\binom{C}{2}$ relevant pairwise combinations. The test statistic we apply is the square of the difference between the local correlation estimates over the grid $\M x_{ij}$. The test variable can thus be defined as follows,
\begin{equation*}
  D_1^*(k,l) =
    \begin{cases}
        \frac{1}{n^2}\sum\limits_{i=1}^{n} \sum\limits_{j=1}^{n} \left[ \hat{\M \rho}_{k}^*(x_i, y_j) - \hat{\M \rho}_{l}^*(x_i, y_j) \right]^2  w(x_i, y_j) & \text{for} \quad k>l \\
         0 & \text{otherwise} \\
    \end{cases}       
\end{equation*}
where $k,l = 1, \ldots C$ and $w$ is a weight function to screen off parts of the local correlation or to concentrate on a certain region. Note that this does not imply disregarding any of the observations, but we choose the weight function such that the distance between the gridpoints and the observations is not too large, i.e. we avoid using an estimated local correlation in a gridpoint far away from any observations. By repeated resampling, $D_1^*(k,l)$ is computed for these resamples and its distribution constructed (i.e. the distribution under $H_0$). From the observations, $\{\V R^c_1, \V R^c_2 \ldots \V R^c_T \}$, calculate $\hat{\M \rho}_{1}(x_i, y_j),  \hat{\M \rho}_{2}(x_i, y_j), \ldots, \hat{\M \rho}_{c}(x_i, y_j)$ and the test statisic,
\begin{equation*}
 D_1(k,l) = \frac{1}{n^2} \sum\limits_{i=1}^n \sum\limits_{j=i}^n \left[ \hat{\M \rho}_{k}(x_i, y_j) -  \hat{\M \rho}_{l}(x_i, y_j) \right]^2  w(x_i, y_j). 
\end{equation*}
The p-value in terms of the $D_1$ distribution is found, and implies a rejection of $H_0$ if it is below a chosen significant level $\alpha$. \Cref{study_error_significance_level} describes an example of this bootstrap test in more detail.
If $C>2$ the statistical analysis involves multiple simultaneous statistical tests, i.e. we face a multiple comparison test problem with $\binom{C}{2}$ pairwise combinations. With large $C$, the number of pairwise tests that we have to perform to confirm/reject the hypothesis increases and with that also the probability of observing rare events, or type I errors. As a consequence, the likelihood of incorrectly rejecting the null hypothesis increases, which we have to adjust for. A classical, but conservative way of dealing with multiple comparison test problem have been to apply the Bonferroni correction. To obtain the Bonferroni corrected/adjusted p-value we divide the original significance level $\alpha$ by the number of tests, $\binom{C}{2}$. Thus, with the Bonferroni correction, we reject the null hypothesis for each pairwise test if the p-value is smaller than $\hat{\alpha_{k,l}} = \alpha_{k,l} / \binom{C}{2}$. There are other, less conservative corrections that can be applied, e.g. Duncan's multiple range test \citep{duncan1955multiple}, Benjamini–Hochberg procedure \citep{benjamini1995controlling} or Holm–Bonferroni method \citep{holm1979simple}). Note, that with $C=2$, there is only one combination to test, and no adjustments are required. In the empirical analysis in \Cref{Empirical_analysis}, we have situations where $C>2$, and thus the p-value needs to be adjusted. 

\subsection{GARCH filtering and bandwidth selection}\label{GARCH_filtering}

LGC estimation requires that the pairs $[R_{1t},R_{2t}], t = 1, ..., T$ are independent and identically distributed (see \cite{tjostheim2013local, bere:tjos:2014}). This is not always realistic, and especially the volatility may exhibit dependence in time. In this paper, we thus apply a GARCH(1,1) filtration to come closer to this assumption, (see \citet{bollerslev1992arch}). In the analysis presented in \Cref{Empirical_analysis}, we filtrate the returns with a GARCH(1,1) model with a student t-distribution. This is also consistent with the approach of e.g. \citet{forbes2002no} and \citet{stov:tjos:2014} in their study of contagion. This, to a sufficient degree levitates the time dependence in the data, and makes it more suitable for the proposed dependency test.   

In \Cref{simulation_studies} we perform simulation studies to check the finite sample performance of the bootstrap test, however, we restrict ourselves for computational reasons to two regimes. We will look at both the error in significance level as well as the power of the test. As described in Section 2, the local Gaussian correlation estimator, $\hat{\rho}(\bm{r})$ depends on two smoothing devices, the bandwidths $\V b = (b_1,b_2)$, and to a lesser degree the kernel used. In the simulations and the empirical analysis we use the Gaussian kernel, and choose the bandwidths using a simple rule of thumb — the global standard deviation times a constant equal to 1.1. This approach gives reasonable results, for instance used in \citet{stov:tjos:huft:2014}, see also \citet{tjostheim2013local} and \citet{bere:tjos:2014} for further discussions regarding bandwidth selection. \Cref{BCNN_decision_alg} presents and summarizes the necessary steps to properly perform the test for asymmetric dependence across regimes, see \cref{Appendix_C}.
 
\section{Simulation studies}\label{simulation_studies}

For the first two simulation set-ups, that is the study of significance level and the power study, we simulate observations with known underlying distributions and dependencies. In the last study, we still simulate from two different distributions, however, we now also classify the observations with a HMM into two regimes, and investigate the power of the test both with the true and predicted observations. A HMM will misclassify some observations, and the purpose of this study is to evaluate how this misclassification impacts the power of the test.

\subsection{Study of error in significance level}\label{study_error_significance_level}

The simulation study for examining the significance level of the proposed test is as follows. The same data generating process (DGP) is used for both regime 1 and 2, hence $H_0$ is true. We use six different DGPs; for every DGP we use two Gaussian marginal distributions, each with a mean equal to zero and a standard deviation equal to four, but with a different copula. The six different copulas are; a Clayton copula with parameter $\theta = 1$, a Clayton copula with parameter $\theta = 2$, a Gaussian copula with parameter $\rho = 0.3$, a Gaussian copula with parameter $\rho = -0.5$, a Gumbel copula with parameter $\theta = 2$ and finally, a Gumbel copula with parameter $\theta = 3$. Hence this set-up closely resembles the set-up in \cite{stov:tjos:huft:2014}, except for the Gaussian copula case with a negative parameter.   
 
The first two models and model 4 are typical models for bivariate equity returns, see e.g. \cite{Okimoto2008}. The negative dependence model 3 corresponds to mimic the often negative relationship observed between bond and equity returns, while models 5 and 6 are included in order to examine how the proposed test behaves in a right-tail dependent environment. 

We generate $M=1000$ independent sets of data from the six models, where each data set is on the form $\{d_1,...,d_T\}$. We set $T=400$ and let regime 1 consists of 300 observations, while regime 2 consists of 100 observations, since usually in practice, one of the regimes will represent a more volatile/bear market period that usually will be shorter than the other regime, representing normal market conditions. Note that for simplicity, we do not perform step 1 in the step-wise procedure above, i.e. we do not fit the regime-switching model to the observations, as we prefer examining the level property of the test under no uncertainty regarding the classification of the observations into two regimes.   

For each model, and for a given set of data, the test statistic $D_1$ is calculated. Bootstrap tests of nominal level 0.01, 0.05 and 0.l0 are conducted based on $B=1000$ bootstrap samples from each of the given data sets, as described above. The null hypothesis is rejected if the proportion of bootstrap statistics exceeding $D_1$ is less than or equal to the appropriate nominal level. Note that the same simulations were used to check all three nominal levels. 

The weight function is in this case chosen such that it corresponds to the range based on the thresholds defined by the $5\%$ lower and $95\%$ upper percentiles, i.e. the weight equals to 1 in this interval and zero outside. Hence, it can vary across the 1000 simulated data sets.

The empirical significance level of the test is reported in Table \ref{tab:error_significant_level}, and the results show that the empirical level of the bootstrap test is consistently close to the nominal level for all models. 

\begin{center}
\begin{table}[H]
\caption{Empirical level of the bootstrap test in the Monte Carlo study. The models correspond to the DGP in both regimes. Each table entry is based on 1000 replications, each with 400 observations.} \label{tab:error_significant_level}
\centering
\begin{tabular}{llll}
\toprule
\multirow{2}{*}{Model} & \multicolumn{3}{r}{Nominal level ($\alpha$)} \\
\cmidrule(r){2-4}
   &  0.01  & 0.05 &  0.1  \\
\midrule
1. Clayton copula, $\theta$ = 1      & 0.017  & 0.056 & 0.102  \\
2. Clayton copula, $\theta$ = 2      & 0.012  & 0.059 & 0.116  \\
3. Gaussian copula, $\rho$ = -0.5    & 0.011  & 0.045 & 0.094   \\
4. Gaussian copula, $\rho$ =  0.3    & 0.007  & 0.058 & 0.116   \\
5. Gumbel copula, $\theta$ = 2    & 0.019  & 0.063 & 0.110    \\
6. Gumbel copula, $\theta$ = 3    & 0.011  & 0.054 & 0.102    \\
\bottomrule
\end{tabular}
\end{table}
\end{center}

\subsection{Study of power}\label{Study of power}

The same setup as for the study of level is used for the study of power, that is, the number of gridpoints and the weight function is similar. The main difference is that the data generating process is different. For regime 1 we use Gaussian copula with $\rho = 0.5$ and two Gaussian marginals both with mean equal to one and standard deviation equal to four. For the second regime, we use six different models. That is, the DGP is still two Gaussian marginal distributions with mean zero and standard deviation equal to four, however, we apply different copulas, as listed in \Cref{tab:study_of_power}. For all the scenarios the $H_1$ hypothesis is true.

In practice, we generate $B=1000$ independent sets of data for each model, each with $T=400$ replications. For the first regime we use 300 samples, while for the second regime we sample 100 replications. This is a similar setup as for the level study. The empirical power is calculated for the three nominal levels $0.01, 0.05$ and $0.1$. The reported power for the different models and nominal levels are reported in \Cref{tab:study_of_power}. In most of the cases examined, the power is acceptable. In particular, for cases 3 and 6 the power is excellent.   

\begin{table}[H]
\caption{Empirical power (times 100) of the bootstrap test in the Monte Carlo study. The models under $H_1$ correspond to the DGP in two different regimes. The DGP in the first regime is a Gaussian copula with $\rho = 0.5$ and with Gaussian marginals. Each table entry is based on 1000 replications, each with 400 observations.}\label{tab:study_of_power}
\centering
\begin{tabular}{llll}
\toprule
\multirow{2}{*}{Model under $H_1$} & \multicolumn{3}{r}{Nominal level ($\alpha)$} \\
\cmidrule(r){2-4}
   & 0.01  &  0.05 &  0.1  \\
\midrule
1. Clayton copula, $\theta$ = 2      & 31.9  & 68 & 82.2  \\
2. Clayton copula, $\theta$ = 3      & 82.1  & 97.8 & 99.7  \\
3. Gaussian copula, $\rho$ = -0.5    & 100  & 100 & 100    \\
4. Gaussian copula, $\rho$ = 0.8    & 73.6  & 93.5 & 96.8    \\
5. Gumbel copula, $\theta$ = 2    & 24.1  & 53.7 & 68.9    \\
6. Gumbel copula, $\theta$ = 3    & 96.3  & 99.4 & 99.9    \\
\bottomrule
\end{tabular}
\end{table}

However, from the \Cref{tab:study_of_power}, we observe that the power of case 1, with Clayton copula, with $\theta=2$, and case 5, the Gumbel copula with $\theta=2$, has much lower power than the other models. The global correlation of these models are approximately $\rho=0.68$ and $\rho=0.70$, respectively. Both models have a correlation that is similar to the model that one is comparing against under $H_0$, hence the power naturally decreases. However, there are ways to improve the power in these cases. For instance, a refinement of the grid for calculation of the LGC could be a possible approach to achieve better power. \cite{stov:tjos:huft:2014} also experienced decreased power for the same models in their test for contagion. The choice of grid size and the fact that we test on the entire grid, and not only on the diagonal, are possible causes that can explain the differences observed between our results and the results in \cite{stov:tjos:huft:2014}. To perform the test on a subset, e.g. only focusing on the lower tail, would certainly improve the power.

Overall, based on the results from both the level and power study, we conclude that the proposed bootstrap test performs as expected for these experiments. The test shows good level and power properties, under the assumption of no misclassification of the observations into the two regimes. The results furthermore indicate that the test is valid.

\subsection{Study of power with HMM classification}

We study how the proposed tests perform when we use a Gaussian multivariate HMM to classify the observations into two regimes. In principle, any conditional distribution could be used, but as we are mainly interested in the difference in dependence structures across regimes, we use the Gaussian distribution for convenience. By using a HMM, we introduce errors due to misclassification, hence we want to assess and compare how the power of the test is affected when introducing potentially wrong regime classifications. To examine this, we design a simulation study where we use a DGP with a Gaussian copula with $\rho=0.5$, with Gaussian marginals with $\mu = 0$ and $\sigma = 3$ for the first regime, while the second regime consists of a Clayton copula with $\theta=3$ with Gaussian marginals with $\mu = 0$ and $\sigma = 5$. 

\begin{figure}[!ht]
    \centering
    \includegraphics[width=\textwidth]{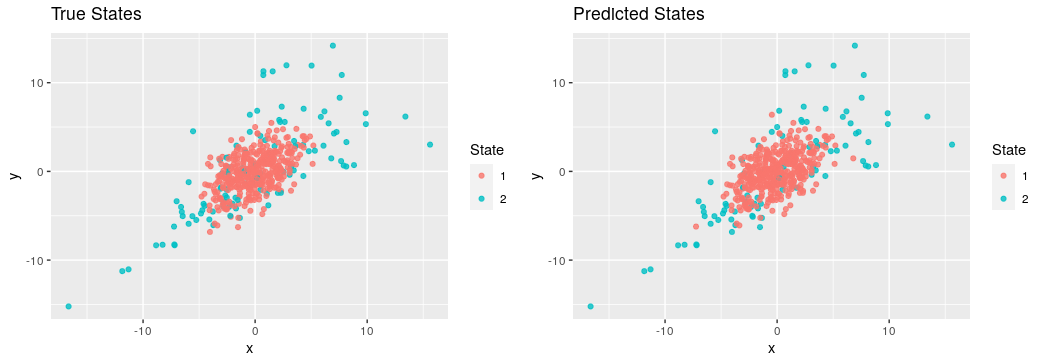}
    \caption{True vs. predicted regimes by a Gaussian HMM for one of the simulated data sets.}
    \label{fig:True_vs_predicted_states}
\end{figure}

We generate 500 data sets each with 500 observations. For each of the 500 data sets we fit a bi-variate Gaussian HMM with TMB in line with \citep{bacri2022gentle}. Furthermore, we use the Backward-Forward algorithm \citep{zucchini} to predict each observation into one of the two regimes for every the data sets. \Cref{fig:True_vs_predicted_states} shows one of the simulated data sets, where the left plot presents the observations with the correct regimes and the right plot, shows how the observations are classified by the HMM into the two regimes. The overall classification accuracy is approximately 79 $\%$ if we assess the classification on all the 500 data sets, see \cref{confusion_matrix}. From \Cref{fig:True_vs_predicted_states} we observe that the model is quite good to identify the correct regime in the observations in both tails. While, in the center of the distribution, the mean, variance and dependency is quite similar, and the HMM struggles somewhat more to classify the observations correctly. 

The confusion matrix is given by
\begin{equation}
    \large
\kbordermatrix{
& \text{Predicted regime 1} & \text{Predicted regime 2}\\
    \text{True regime 1}  & 68.5 \% & 6.8 \%   \\
     \text{True regime 2} & 14.2 \%   & 10.5\%
  },
\end{equation}\label{confusion_matrix}

showing that the HMM models across all 500 simulated data sets have relatively good prediction capabilities when it comes to correctly classifying the observations. 

\begin{figure}
    \centering
    \includegraphics[width=\textwidth]{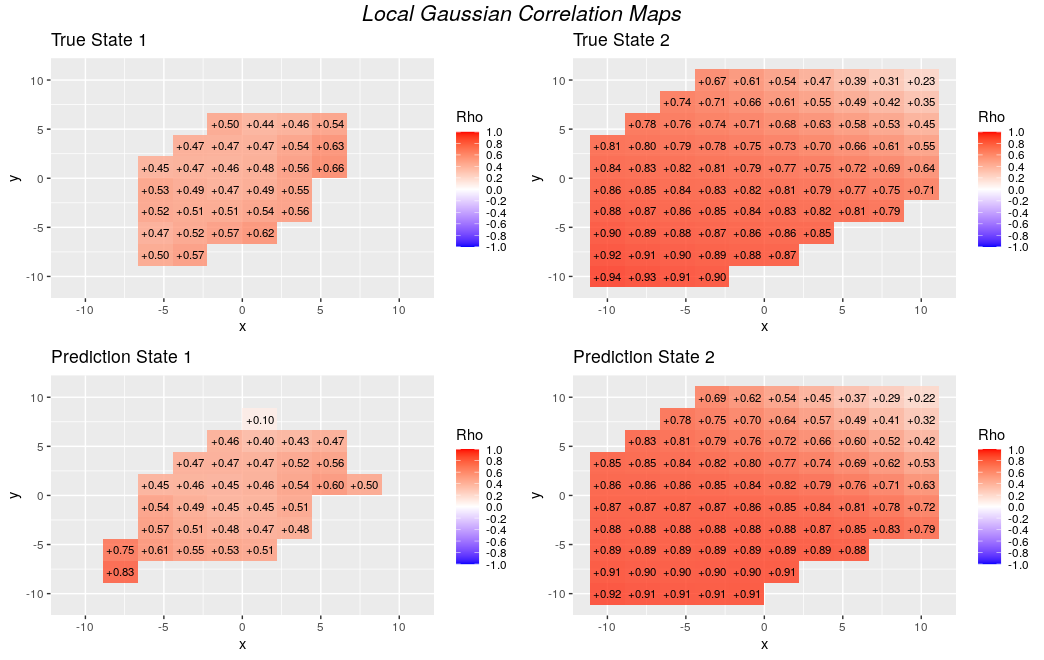}
    \caption{LGC-map of true vs. predicted regimes.}
    \label{fig:LGC_True_vs_predicted_states}
\end{figure}

After the classification of the observations into two regimes by the fitted HMM for each of the 500 simulated data sets, we calculate the LGC and apply the asymmetric dependency test similarly as in \Cref{Study of power} for each realization. For comparison, we perform the test twice for each data set, that is, on the observations arising from the predicted regimes and on the observations from the true regimes.  \Cref{fig:LGC_True_vs_predicted_states} shows the LGC-map of one of the data sets generated, using the observations from the true regimes on the left plot, while using the observations from the predicted regimes on the right plot. We observe a significant difference in the estimated LGC map when comparing regime 1 and regime 2 for both models, as expected from the DGP. However, we observe only minor differences in the LGC maps when comparing the map based on the observations using the true regimes versus the predicted regimes. This is indeed a positive finding, as it implies that the fitting of the HMM and the corresponding classification of the observations into the two regimes, are only marginally impacting the estimated LGC. 

The power of the test from the two scenarios, both based on the true and on the predicted observations are shown in \Cref{tab:power_test_with_HMM}.      

\begin{table}[H]
\caption{Empirical power (times 100) of the dependency bootstrap test in the Monte Carlo study.}\label{tab:power_test_with_HMM}
\centering
\begin{tabular}{ccllll}
\hline
\multirow{2}{*}{\centering{Model under $H_1$}} & \multirow{2}{*}{regime} &\multicolumn{3}{c}{Nominal level ($\alpha)$} \\
\cmidrule(r){3-5}
& & 0.01 & 0.05 & 0.1 \\
\hline
1. Clayton copula, $\theta$ = 3 & True & 96.2 & 99.4 & 99.8 \\
2. Clayton copula, $\theta$ = 3 & Predicted & 64.6 & 81.8 & 87.2  \\
\hline
\end{tabular}
\end{table}

The 500 HMMs have approximately $21 \%$ of the observations misclassified. Due to this misclassification we observe a small degeneration of power in the cases with predicted regimes compared with the cases using the true regmies. Dependent on the nominal level, the degeneration of the power is $32.8\%$, $17.7\%$, and $12.6\%$ for nominal levels $0.01$, $0.05$ and $0.1$, respectively. However, the power is still acceptable, and we conclude that the test is performing reasonably well also in the case where regime classification is performed.   

\newpage

\section{Empirical analysis}\label{Empirical_analysis}

In this section we illustrate our approach on two real data sets. We want to assess and find differences in LGC between two financial time series. For stock indices, e.g. S\&P500 vs. FTSE100, it is observed that in a falling market, there is a different dependence structure than in more stable periods (see e.g. \cite{Okimoto2008}). To identify stable versus more unstable regimes we can use a HMM, as outlined in \Cref{sec:HMMs}. Our objective is, given a set of regimes obtained from a classification with a HMM, to identify whether or not the LGC in the regimes are significantly different using the proposed bootstrap test. In the first empirical analysis we use well known financial time series of the US stock index, the S\&P500, and the UK index, FTSE100. Secondly, we investigate weekly data from S\&P500 and US Bonds with 10 year maturity (BMUS10Y). 

\subsection{S\&P500 and FTSE100}

The combination of the US index S\&P500 and UK index FTSE100 have been used in numerous analyses e.g. \citep{longin:2001, Okimoto2008, stov:tjos:huft:2014}. The time series consist of 9006 daily observations from $1987/07/09$ to $2022/11/01$. In \Cref{tab:2_state_SP500_vs_FTSE100} an overview of descriptive statistics for both the S\&P500 and FTSE100 returns is given. The data exhibit non-normality,
as is seen from the skewness and kurtosis coefficients. Also, the Jarque–Bera test rejects normality for all series. See \Cref{Appendix_A} for details on the parameters of the HMM fitting. Regime 2 is more volatile than regime 1, as we can see from the statistics in \Cref{tab:2_state_SP500_vs_FTSE100}, in particular the variance and kurtosis is much larger in regime 2 than regime 1. 
\begin{table}[H]
\centering
\begingroup\footnotesize
\caption{Descriptive statistics for a 2 regime classification of daily returns from S\&P500 and FTSE100.}
\begin{tabular}{llrrrrrrrrrr}
 \textbf{Variable} & \textbf{Levels} & $\mathbf{n}$ & $\mathbf{\bar{x}}$ & $\mathbf{\widetilde{x}}$ & \textbf{Min} & \textbf{Max} & \textbf{IQR} & $\textbf{Variance}$ & $\textbf{Skewness}$ & $\textbf{Kurtosis}$ & $\textbf{Jarque-Bera}$ \\ 
  \hline
S\&P500 & Regime 1 & 7038 &  0.1 &  0.0 &  -2.6 &  2.8 & 0.8 & 0.5 &  0.0 &  3.7 &    163.0 \\ 
   & Regime 2 & 2109 & -0.1 &  0.0 & -22.9 & 11.0 & 2.5 & 4.3 & -0.8 & 12.6 &   8232.7 \\ 
   \hline
 & all & 9147 &  0.0 &  0.0 & -22.9 & 11.0 & 1.0 & 1.4 & -1.2 & 29.3 & 266689.3 \\ 
   \hline
S\&P500 GF & Regime 1 & 7038 &  0.0 &  0.0 &  -4.4 &  3.7 & 1.0 & 0.8 & -0.2 &  4.2 &    504.5 \\ 
   & Regime 2 & 2109 & -0.2 & -0.1 & -10.4 &  4.1 & 1.6 & 1.7 & -1.0 &  7.4 &   2050.7 \\ 
   \hline
 & all & 9147 & -0.1 &  0.0 & -10.4 &  4.1 & 1.1 & 1.0 & -0.7 &  7.5 &   8442.7 \\ 
   \hline
FTSE100 & Regime 1 & 7038 &  0.1 &  0.0 &  -2.6 &  2.8 & 0.9 & 0.5 & -0.1 &  3.4 &     59.8 \\ 
   & Regime 2 & 2109 & -0.1 & -0.1 & -13.0 &  9.4 & 2.2 & 3.5 & -0.3 &  6.7 &   1250.6 \\ 
   \hline
 & all & 9147 &  0.0 &  0.0 & -13.0 &  9.4 & 1.1 & 1.2 & -0.6 & 13.7 &  43760.0 \\ 
   \hline
FTSE100 GF & Regime 1 & 7038 &  0.1 &  0.0 &  -2.6 &  2.8 & 0.9 & 0.5 & -0.1 &  3.4 &     59.8 \\ 
   & Regime 2 & 2109 & -0.1 & -0.1 & -13.0 &  9.4 & 2.2 & 3.5 & -0.3 &  6.7 &   1250.6 \\ 
   \hline
 & all & 9147 &  0.0 &  0.0 & -13.0 &  9.4 & 1.1 & 1.2 & -0.6 & 13.7 &  43760.0 \\ 
   \hline
\end{tabular}
\label{tab:2_state_SP500_vs_FTSE100}
\endgroup
\end{table}

\begin{figure}[!ht]
    \centering
    \includegraphics[width=\textwidth]{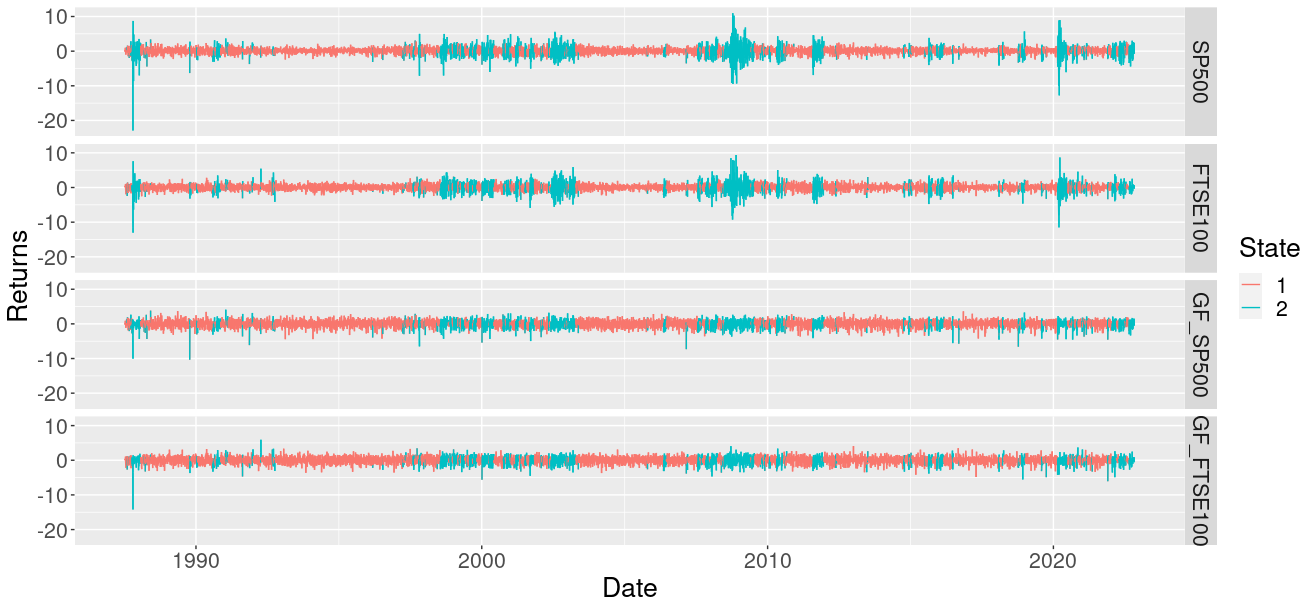}
    \caption{Log-returns and GARCH filtrated log-returns of S\&P500 and FTSE100 with classification into two regimes using a Gaussian hidden Markov model.}
    \label{fig:FTSE100_TS}
\end{figure}

\begin{figure}[!ht]
    \centering
    \includegraphics[width=\textwidth]{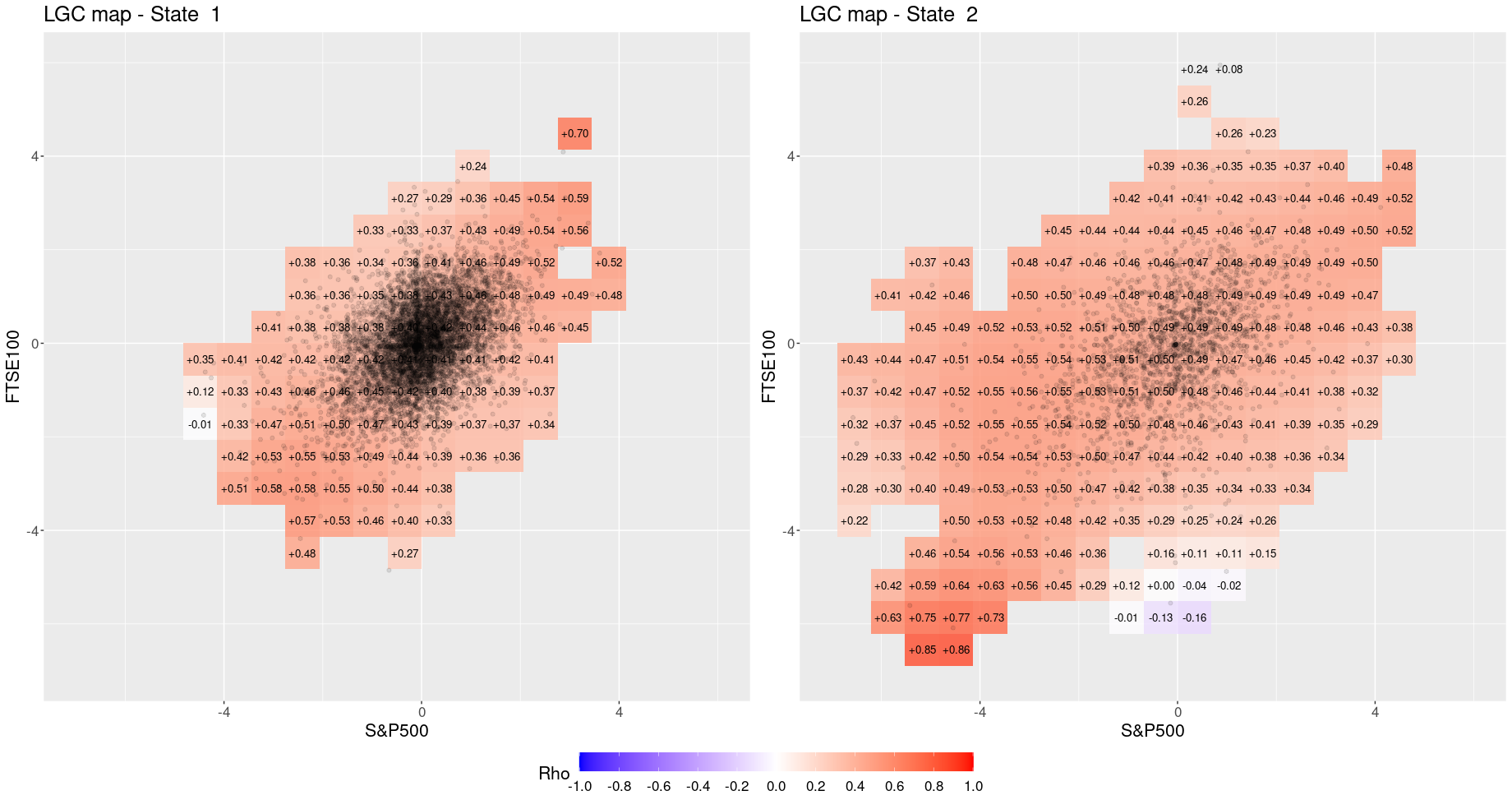}
    \caption{LGC regime 1 and 2 for S\&P500 and FTSE100 daily data.}
    \label{fig:SP500_vs_FTSE_LGC_map}
\end{figure}

The two upper plots in \Cref{fig:FTSE100_TS} shows the log-returns of S\&P500 and FTSE100. Furthermore, the observations have been classified into two regimes (coloured red and green). This classification is the output of the bivariate Gaussian HMM with the assumption that there exists two regimes. Comparing the classification with historical events, it seems that the fitted HMM has the ability to identify crisis periods/bear markets. The observations around the financial crisis of 2007-2009 and the COVID-19 pandemic in 2020, both belongs to regime 2. Furthermore, observations during the financial crash of October 1987, and during the dot-com bubble in the early 2000s, are also classified as regime 2. The majority of the observations are, however, classified as regime 1, or stable/bull market periods. As mentioned, the returns are next filtered by a separate GARCH(1,1) model with a Student-t distribution for each of the time series, and the estimated parameters are given in \Cref{tab:GARC_sp500_daily} and \Cref{tab:GARCH_FTSE_daily}. The GARCH filtrated time series, that will be used in the statistical test, are presented in the two lower plots of \Cref{fig:FTSE100_TS}. The volatility clustering, and time dependence in the observations are clearly reduced by this filtering. 
The observed LGC for the two regimes is presented in \Cref{fig:SP500_vs_FTSE_LGC_map}. Visual inspection of the LGC map for the two regimes, shows that the variance and local correlation is larger for regime 2 than regime 1. The correlations for regime 1 (stable/bull market) are relatively uniform. In the second regime (bear market) we observe an asymmetric dependency structure with a larger correlation in the lower-left tail of the distribution/LGC-map. That is, in a bear market we have identified a stronger tail dependency than in the bull market. This is in line with other studies using regime switching copulas \citep{Rodriguez2007, Okimoto2008, bens:2018}. The benefit of our approach is that it is much easier to interpret and understand the dependence structures. There is no need to have any assumption on the (bear market) dependency structure beforehand, i.e. by specifying a certain copula. In both the bull and bear market, the dependency structure is directly revealed through the LGC maps.    
Finally, to examine whether the two regimes have different dependency structures, we apply the test outlined in \Cref{sec:test}. The test is performed with 1000 bootstrap replicates, which gives a p-value of 0.001. This means that the null hypothesis is clearly rejected for all reasonable significance levels, and we conclude that the dependency structures of the two regimes are statistically significantly different from each other. 

\subsection{S\&P500 and US Bonds}

The dependence relationships between different asset classes, as stocks, bonds and commodities, have been widely studied, see e.g. \cite{dajcman:2012}, \cite{aslanidis2012} and \cite{jammazi:2015}. The main reason for studying these relationships is that different asset classes typically represent the building blocks of most investment portfolios because of their different risk-return characteristics, and in particular the stocks and bonds linkage is important in this respect. In the next empirical analysis, we thus study the stock-bond relationship with our proposed procedure, and perform equality tests across different regimes. We further align our findings with the current knowledge of the stock-bond relationship throughout the analysis. It is well-known that there is substantial time variation in the co-movement. Until the mid-1990s the US stock-bond correlation was strongly positive, and then changed to a negative correlation by the early 2000s and onwards. Furthermore, some authors have also used a copula approach, for instance \cite{jammazi:2015}. They document a lack of tail dependence in the stock-bond relation, which suggests that stock and bond markets do not tend to boom or crash together. Further, the dependence seems not especially strong during extreme market conditions, but rather that it is present most of the time.

\begin{figure}[!ht]
    \centering
    \includegraphics[width=\textwidth]{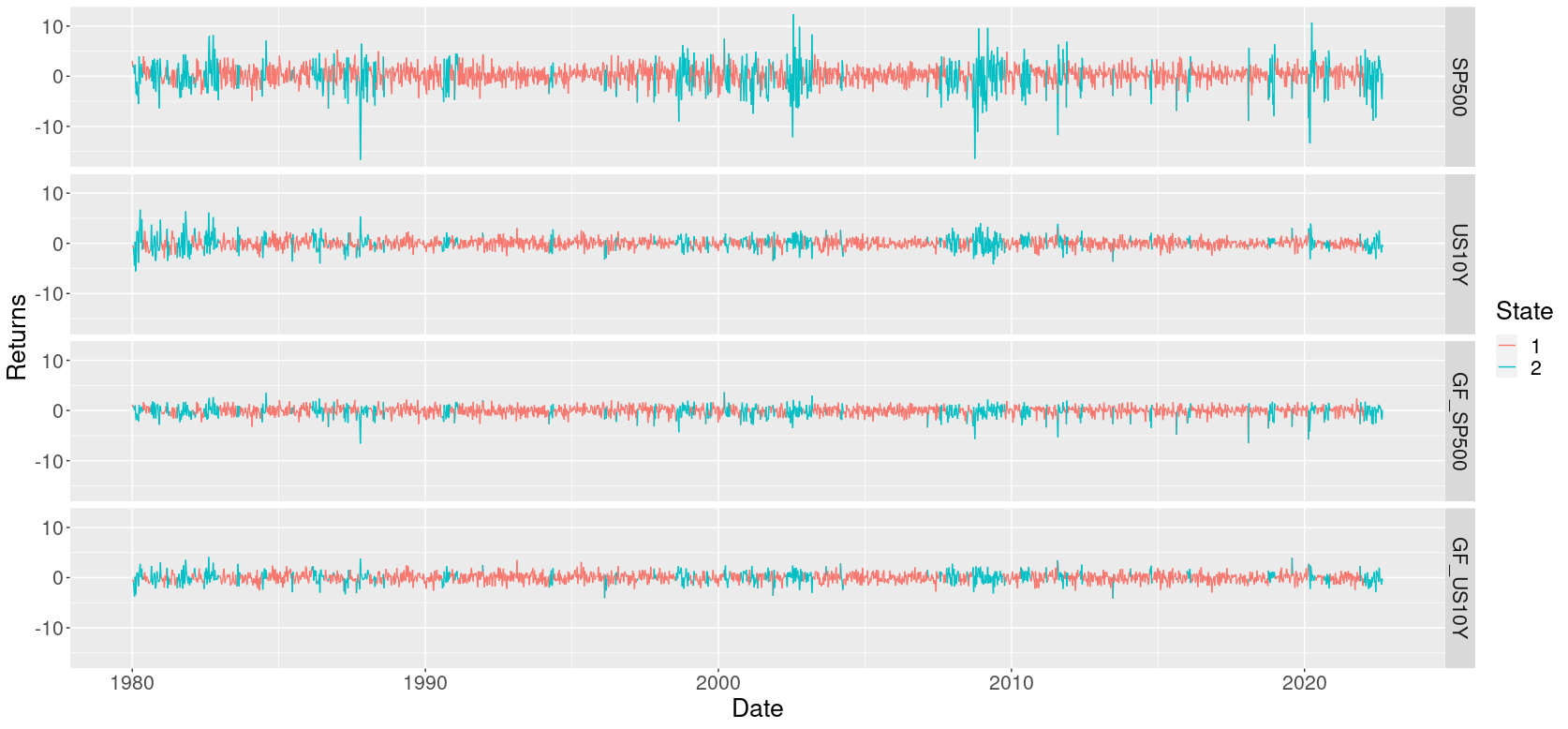}
    \caption{Log-return of S\&P500 and BMUS10Y with classification with a HMM with two regimes.}
    \label{fig:SP500_vs_BMUS_log_return_two_states}
\end{figure}

We use weekly log-returns of the S\&P500 and US Bonds with 10 year maturity (BMUS10Y) from $1980/01/02$ to $2022/08/31$. In the first part of this section we assume that there exists two regimes, i.e. a bull and bear market for these observations. The results of the HMM classification are visually presented in \Cref{fig:SP500_vs_BMUS_log_return_two_states} along with descriptive statistics in Table \ref{tab:DS_US10Y_2_states}. The optimized HMM model parameters are presented in \Cref{Appendix_A}. We perform the classification on raw returns, however, for estimation of the LGC map, we use the GARCH(1,1)-filtrated data for the same reasons mentioned for the empirical analysis of stock indices, see \Cref{tab:GARCH_SP500_weekly_data} for the estimated parameters. \Cref{tab:DS_US10Y_2_states} shows that there are in total 2220 pair of observations, where 1664 are classified as bull market, or regime 1, and 556 are classified as bear market or, regime 2. The bear market has less observations, higher variance, IQR, minimum and maximum values. We observe higher Jarque-Berra values for the bear compared to the bull market, i.e. the bear markets are less Gaussian than the bull markets. We also observe that the GARCH filtrated data have a mean closer to zero and a lower IQR. 

\begin{table}[H]
\centering
\begingroup\footnotesize
\caption{Descriptive statistics for 2 regime classification of weekly returns from S\&P500 and BMUS10Y.} 
\begin{tabular}{llrrrrrrrrrr}
 \textbf{Variable} & \textbf{Levels} & $\mathbf{n}$ & $\mathbf{\bar{x}}$ & $\mathbf{\widetilde{x}}$ & \textbf{Min} & \textbf{Max} & \textbf{IQR} & $\textbf{Variance}$ & $\textbf{Skewness}$ & $\textbf{Kurtosis}$ & $\textbf{Jarque-Bera}$ \\ 
  \hline
S\&P500 & Regime 1 & 1664 &  0.4 &  0.4 &  -5.0 &  5.3 & 1.8 &  2.1 & -0.1 & 3.2 &    9.5 \\ 
   & Regime 2 &  556 & -0.4 & -0.4 & -16.7 & 12.4 & 5.3 & 14.4 & -0.4 & 4.2 &   46.6 \\ 
   \hline
 & all & 2220 &  0.2 &  0.3 & -16.7 & 12.4 & 2.4 &  5.3 & -0.9 & 8.8 & 3374.1 \\ 
   \hline
S\&P500.GF & Regime 1 & 1664 &  0.0 &  0.1 &  -3.2 &  2.5 & 1.0 &  0.6 & -0.3 & 3.4 &   41.9 \\ 
   & Regime 2 &  556 & -0.4 & -0.2 &  -6.6 &  3.6 & 2.0 &  2.0 & -0.7 & 4.3 &   86.4 \\ 
   \hline
 & all & 2220 & -0.1 &  0.0 &  -6.6 &  3.6 & 1.1 &  1.0 & -0.9 & 6.3 & 1317.5 \\ 
   \hline
BMUS10Y & Regime 1 & 1664 &  0.0 &  0.0 &  -3.0 &  3.0 & 1.2 &  0.7 &  0.0 & 3.2 &    2.8 \\ 
   & Regime 2 &  556 &  0.1 &  0.1 &  -5.6 &  6.7 & 2.1 &  2.9 &  0.2 & 3.8 &   16.2 \\ 
   \hline
 & all & 2220 &  0.0 &  0.0 &  -5.6 &  6.7 & 1.3 &  1.3 &  0.2 & 5.7 &  710.4 \\ 
   \hline
BMUS10Y.GF & Regime 1 & 1664 &  0.0 &  0.0 &  -3.0 &  3.5 & 1.1 &  0.7 &  0.0 & 3.2 &    2.3 \\ 
   & Regime 2 &  556 &  0.1 &  0.1 &  -4.2 &  4.1 & 1.8 &  1.8 & -0.1 & 3.2 &    2.5 \\ 
   \hline
 & all & 2220 &  0.0 &  0.0 &  -4.2 &  4.1 & 1.3 &  1.0 &  0.0 & 3.8 &   61.4 \\ 
   \hline
\end{tabular}\label{tab:DS_US10Y_2_states}
\endgroup
\end{table}

Stock indices usually 
have positive dependence both in crisis and non-crisis periods, with however a stronger degree of dependence in crisis periods in the tail. S\&P500 time series vs. US10Y have a different dependency structure. Both in the bull and bear market we observe an asymmetric dependency behaviour when we assume there exists 2 regimes. The variability of the bull market LGC map is somewhat lower than for the bear market LGC map, but the same underlying dependency structure is observed. That is, on the diagonal, the local correlations are negative and conversely on the cross-diagonal. As will be examined in \Cref{sec: three_state_classification}, where we will fit models with 3 and more regimes, this asymmetry is due to a shift in the bull market behaviour in the early 2000s, where the dependency shifted from positive to a negative dependency. In essence, this empirical analysis shows that a two regime Gaussian HMM has too few parameters to identify this shift in the bull market dependency structure. This change in bull market dependency was also highlighted by \citep{jammazi:2015} in his paper on time-varying dependence between stock and government bond returns.

\begin{figure}[!ht]
    \centering
    \includegraphics[width=\textwidth]{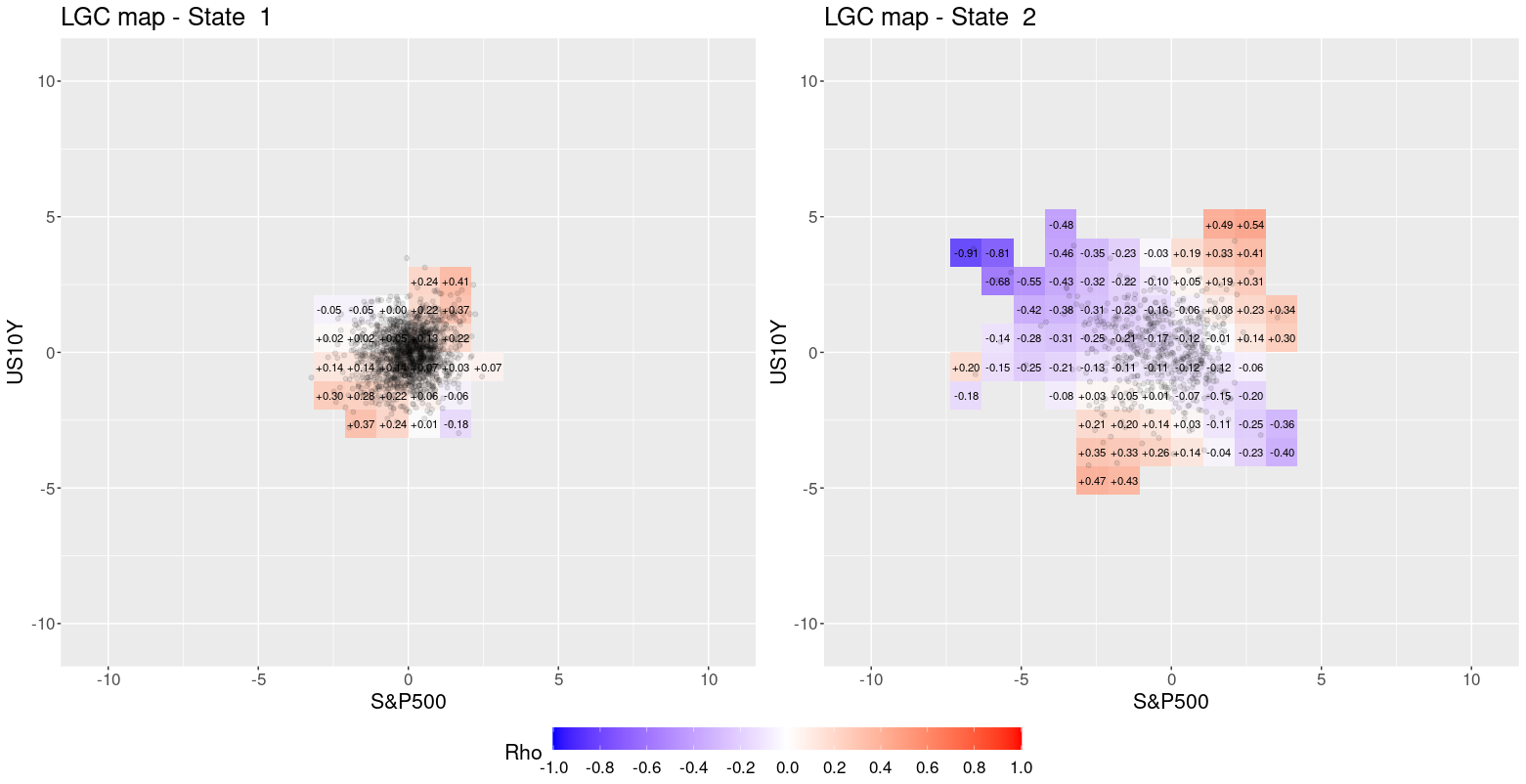}
    \caption{LGC map of S\&P500 and BMUS10Y two regimes.}
    \label{fig:SP500_vs_BMUS_LGC_map_two_states}
\end{figure}
We have performed the dependency test on the classified GARCH-filtrated data and the two corresponding LGC maps. Unsurprisingly, the test shows that the dependence between the two regimes are significantly different, i.e. the null hypothesis is rejected. The p-value calculated with the asymmetric dependence bootstrap test was in fact 0 for this case.  
\newpage

\subsubsection{Model selection}

So far we have performed empirical analysis with two different regimes, in line with previous work e.g. \citet{Okimoto2008}. From \citet{jammazi:2015}, we know there may exist more than one dependency structure for the bull market.  For the S\&P500 and BMUS10Y data we want to examine this further by assessing different models by evaluating the AIC and BIC with several number of regimes. We fit models up to 6 regimes. For each of the models, we calculate the AIC and BIC, and both AIC and BIC are relative estimators of the prediction error and thus can be used for model selection. The main difference between the two criteria are that the BIC penalizes the number of parameters harder. The number of parameters increases with the number of regimes specified for the HMMs. Hence, BIC favours models with fewer regimes than AIC. \Cref{tab:AIC_BIC} shows the AIC and BIC calculated for different HMMs with different number of regimes. The BIC favours a model with 3 regimes whereas the AIC selects a model with 5 regimes.

\begin{table}[!ht]
\centering
\caption{Calculated AIC and BIC dependent on number of regimes.}
\begin{tabular}{ccc}
  \hline
\# of regimes & AIC & BIC \\ 
  \hline
  1 & 9852.45 & 9880.98 \\ 
    2 & 9506.66 & 9575.13 \\ 
    3 & 9376.73 & \textbf{9496.54} \\ 
    4 & 9340.52 & 9523.08 \\ 
    5 & \textbf{9316.37} & 9573.10 \\ 
    6 & 9325.96 & 9668.28 \\ 
   \hline
\end{tabular}\label{tab:AIC_BIC}
\end{table}

\subsubsection{Three-regime model}\label{sec: three_state_classification}

Table \ref{tab:desc_stat_3_states_US10Y} shows descriptive statistics for the three regime model of the time series S\&P500 and US10Y log-returns. Regime 2 contains the most observations with 1115 while regime 1 consists of 743 observations, and regime 3 contains 362 observations. Regime 1 and 2 are less volatile and have lower IQR and smaller maximum and minimum values than regime 2. The GARCH filtrated time series (note that we utilize the same GARCH-model as for the case with two regimes) have a lower mean, IQR and extreme values than the raw log-returns.   

\begin{table}[!ht]
\centering
\begingroup\footnotesize
\caption{Descriptive statistics for 3 regime classification of weekly returns from S\&P500 and BMUS10Y.} 
\label{tab:desc_stat_3_states_US10Y}
\begin{tabular}{llrrrrrrrrrr}
 \textbf{Variable} & \textbf{Levels} & $\mathbf{n}$ & $\mathbf{\bar{x}}$ & $\mathbf{\widetilde{x}}$ & \textbf{Min} & \textbf{Max} & \textbf{IQR} & $\textbf{Variance}$ & $\textbf{Skewness}$ & $\textbf{Kurtosis}$ & $\textbf{Jarque-Bera}$ \\ 
  \hline
S\&P500 & Regime 1 &  743 &  0.2 &  0.3 &  -5.6 &  5.6 & 1.8 &  2.9 & -0.5 & 3.9 &   48.3 \\ 
   & Regime 2 & 1115 &  0.3 &  0.4 &  -6.4 &  5.3 & 2.2 &  3.0 & -0.3 & 3.5 &   27.5 \\ 
   & Regime 3 &  362 & -0.4 & -0.2 & -16.7 & 12.4 & 5.4 & 17.2 & -0.5 & 4.3 &   37.3 \\ 
   \hline
 & all & 2220 &  0.2 &  0.3 & -16.7 & 12.4 & 2.4 &  5.3 & -0.9 & 8.8 & 3374.1 \\ 
   \hline
S\&P500 GF & Regime 1 &  743 & -0.1 &  0.0 &  -3.6 &  2.5 & 0.9 &  0.8 & -0.9 & 4.7 &  176.3 \\ 
   & Regime 2 & 1115 &  0.0 &  0.0 &  -3.3 &  2.2 & 1.1 &  0.8 & -0.4 & 3.3 &   35.0 \\ 
   & Regime 3 &  362 & -0.3 & -0.1 &  -6.6 &  3.6 & 1.8 &  2.1 & -1.0 & 5.6 &  157.7 \\ 
   \hline
 & all & 2220 & -0.1 &  0.0 &  -6.6 &  3.6 & 1.1 &  1.0 & -0.9 & 6.3 & 1317.5 \\ 
   \hline
BMUS10Y & Regime 1 &  743 &  0.1 &  0.1 &  -2.5 &  3.1 & 1.0 &  0.7 &  0.0 & 3.5 &    7.3 \\ 
   & Regime 2 & 1115 & -0.1 & -0.1 &  -3.7 &  3.3 & 1.4 &  1.0 & -0.1 & 3.3 &    4.5 \\ 
   & Regime 3 &  362 &  0.2 &  0.1 &  -5.6 &  6.7 & 2.1 &  3.4 &  0.2 & 3.8 &   14.1 \\ 
   \hline
 & all & 2220 &  0.0 &  0.0 &  -5.6 &  6.7 & 1.3 &  1.3 &  0.2 & 5.7 &  710.4 \\ 
   \hline
BMUS10Y GF & Regime 1 &  743 &  0.1 &  0.1 &  -3.0 &  3.9 & 1.1 &  0.7 &  0.0 & 3.7 &   14.7 \\ 
   & Regime 2 & 1115 & -0.1 & -0.1 &  -4.2 &  3.5 & 1.3 &  0.9 & -0.1 & 3.5 &   13.9 \\ 
   & Regime 3 &  362 &  0.1 &  0.0 &  -3.7 &  4.1 & 1.7 &  1.7 &  0.0 & 3.3 &    1.1 \\ 
   \hline
 & all & 2220 &  0.0 &  0.0 &  -4.2 &  4.1 & 1.3 &  1.0 &  0.0 & 3.8 &   61.4 \\ 
   \hline
\end{tabular}
\endgroup
\end{table}

\begin{figure}[!ht]
    \centering
    \includegraphics[width=\textwidth]{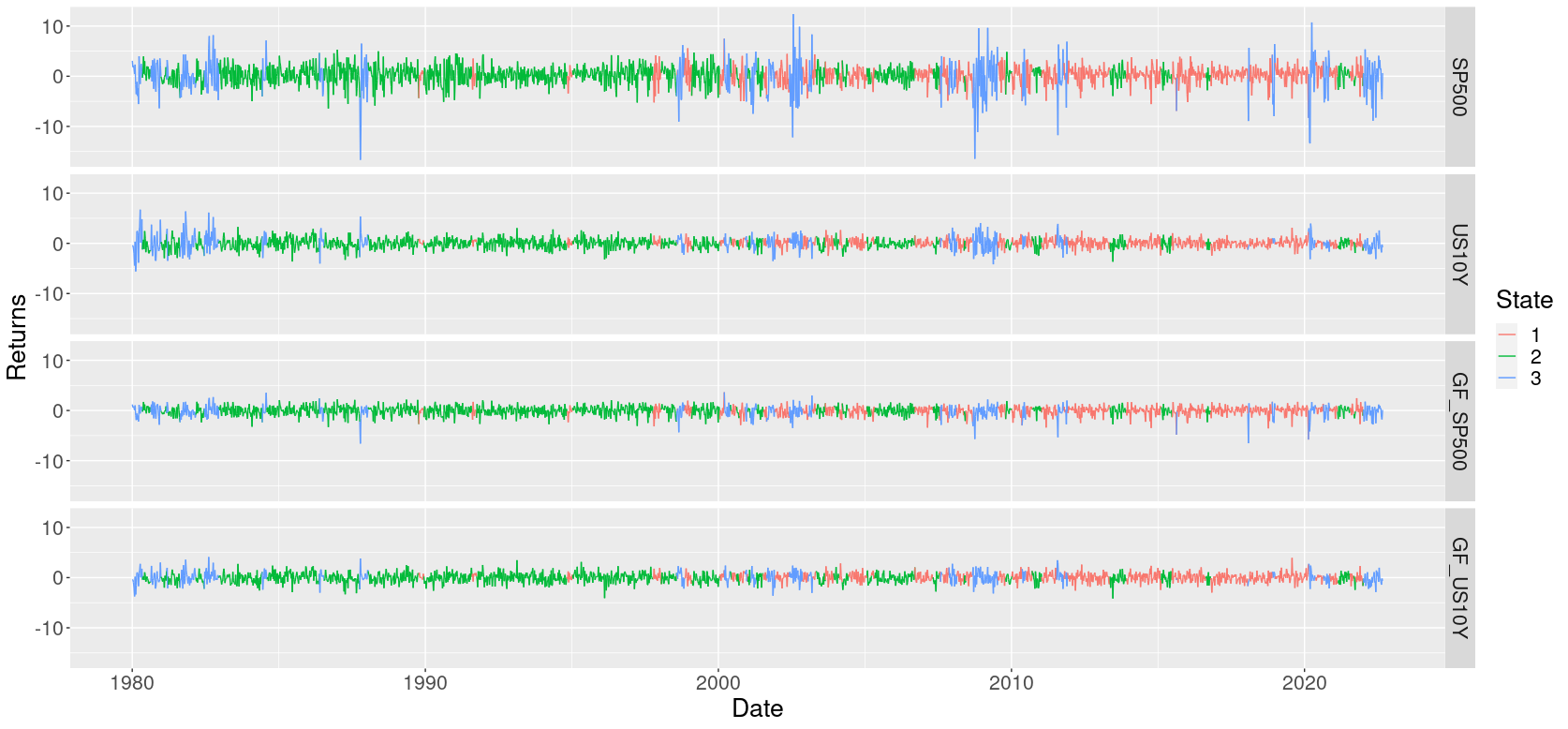}
    \caption{LGC map of S\&P500 and BMUS10Y three regimes.}
    \label{fig:SP500_vs_BMUS_Returns_three_states}
\end{figure}

\begin{figure}[!ht]
    \centering
    \includegraphics[width=\textwidth]{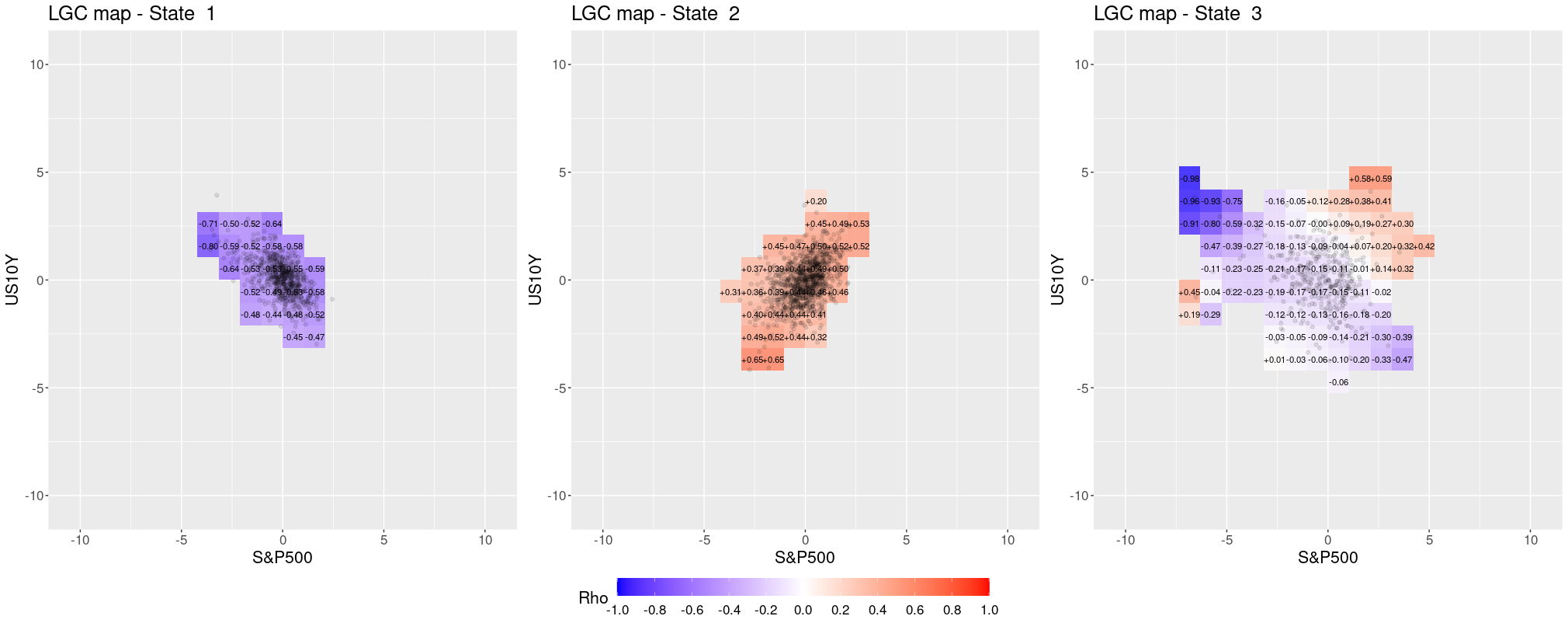}
    \caption{LGC map of S\&P500 and BMUS10Y three regimes.}
    \label{fig:SP500_vs_BMUS_LGC_map_three_states}
\end{figure}

\Cref{fig:SP500_vs_BMUS_Returns_three_states} shows the log-returns and the GARCH filtrated returns. The colours red, green and blue shows the classification for regime 1, 2 and 3, respectively. 
\Cref{fig:SP500_vs_BMUS_LGC_map_three_states} shows the LGC map for observations for the three different regimes. The first regime, has a clear negative correlation, the second regime a clear positive correlation, while the third regime has an asymmetric dependency structure. 
We observe that regime two is mainly present from the 80s to the beginning of the 00s. When examining the corresponding estimated LGC map, we observe a clear positive dependence. From the 00s, and until 2022, the observations in the calmer period (regime 1) have a negative dependence. This change in dependency is well known (see e.g. \cite{jammazi:2015}) between US stock indices and US bonds.

The more volatile period, that is regime 3, is observable throughout the time period under study and seems to mainly correspond to historical recessions or crisis. Where both regime 1 and 2 have a relatively uniform LGC structure, regime 3 has an asymmetric dependency structure. Compared to regime 1, the negative correlation is reduced, and we also observe positive local correlations in some segments. Hence, in crisis periods, the anticipated diversification benefits is clearly reduced when comparing to regime 1.

Comparing the two-regime model with the three-regime model, we observe that the most volatile and turmoil period in both models are relatively equal. In regime 1 of the two-regime model, we see a clear asymmetric dependency structure which is not present in regime 1 and 2 in the three-regime model. Assessing the observations in both models, more or less regime 1 and regime 2 in the three-regime model attracts the same observations as regime 1 in the two-regime model. Due to the underparametrization of the two-regime model, the change in dependency structure is not identified, which in turn is reflected in the LGC map. 

With more than two regimes we need to adjust the p-values because we face the multi-comparison problem. However, as expected by examining the LGC maps (\Cref{fig:SP500_vs_BMUS_LGC_map_three_states}), the differences in the LGCs are large, and the asymmetric dependency test rejects the null hypothesis in all pairwise tests. In other words, the p-value is 0 for all of the three relevant pairwise tests.


\subsubsection{Five-regime model}\label{sec:AIC_five_state_US10Y} 

\Cref{tab:5_state_US10Y} shows descriptive statistics for the 5 different regimes in the HMM classification. Regime 5 has relatively few observations, and the estimated LGCs should thus be viewed with care. 

\begin{figure}[h!]
    \centering
    \includegraphics[width=\textwidth]{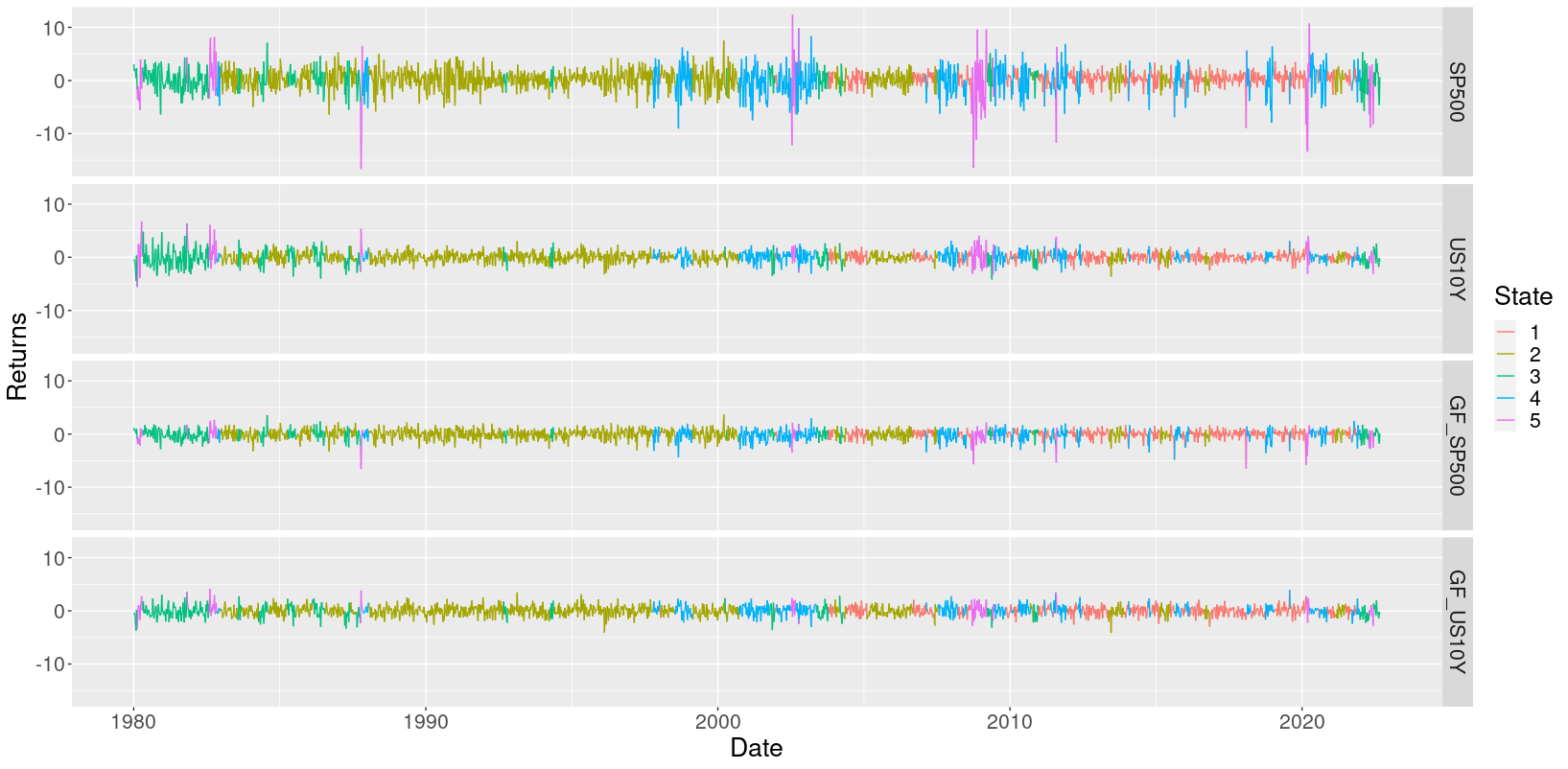}
    \caption{TS Classification of S\&P500 and BMUS10Y five regimes.}
    \label{fig:TS_5_states}
\end{figure}

\begin{figure}[h!]
    \centering
    \includegraphics[width=\textwidth]{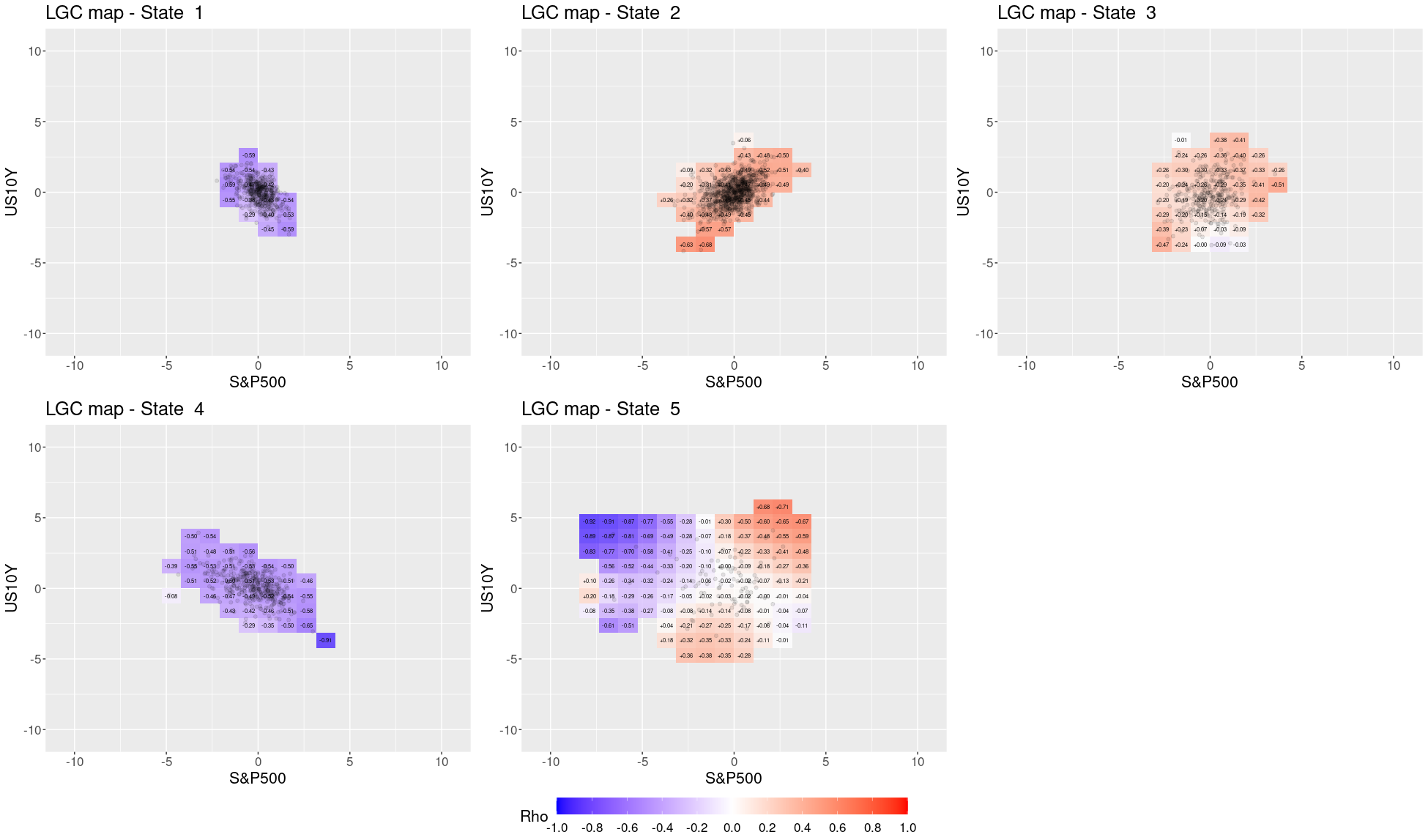}
    \caption{LGC map of S\&P500 and BMUS10Y five regimes.}
    \label{fig:LGC_five_States}
\end{figure}

\begin{table}[!ht]
\centering
\begingroup\footnotesize
\caption{Descriptive statistics for 5 regime classification of weekly returns from S\&P500 and BMUS10Y.} 
\label{tab:5_state_US10Y}
\begin{tabular}{llrrrrrrrrrr}
 \textbf{Variable} & \textbf{Levels} & $\mathbf{n}$ & $\mathbf{\bar{x}}$ & $\mathbf{\widetilde{x}}$ & \textbf{Min} & \textbf{Max} & \textbf{IQR} & $\textbf{Variance}$ & $\textbf{Skewness}$ & $\textbf{Kurtosis}$ & $\textbf{Jarque-Bera}$ \\ 
  \hline
S\&P500 & Regime 1 &  480 &  0.4 &  0.5 &  -2.7 &  3.6 & 1.4 &  1.2 & -0.2 & 3.3 &    5.3 \\ 
   & Regime 2 &  952 &  0.3 &  0.4 &  -6.4 &  7.5 & 2.2 &  3.1 & -0.2 & 3.7 &   24.0 \\ 
   & Regime 3 &  301 &  0.2 &  0.3 &  -6.4 &  7.1 & 2.7 &  4.5 & -0.2 & 3.1 &    2.1 \\ 
   & Regime 4 &  398 & -0.2 & -0.2 &  -9.0 &  8.3 & 4.3 &  8.7 & -0.1 & 2.6 &    3.5 \\ 
   & Regime 5 &   89 & -1.1 & -0.3 & -16.7 & 12.4 & 7.9 & 37.2 & -0.3 & 2.9 &    1.2 \\ 
   \hline
 & all & 2220 &  0.2 &  0.3 & -16.7 & 12.4 & 2.4 &  5.3 & -0.9 & 8.8 & 3374.1 \\ 
   \hline
S\&P500 GF & Regime 1 &  480 &  0.1 &  0.1 &  -2.2 &  1.7 & 0.8 &  0.4 & -0.4 & 3.6 &   19.4 \\ 
   & Regime 2 &  952 &  0.0 &  0.0 &  -3.3 &  3.6 & 1.1 &  0.8 & -0.3 & 3.5 &   29.4 \\ 
   & Regime 3 &  301 & -0.1 &  0.0 &  -2.8 &  3.5 & 1.2 &  1.0 & -0.2 & 3.2 &    1.8 \\ 
   & Regime 4 &  398 & -0.3 & -0.2 &  -4.8 &  3.0 & 1.5 &  1.4 & -0.6 & 3.5 &   27.5 \\ 
   & Regime 5 &   89 & -0.6 & -0.1 &  -6.6 &  2.7 & 2.3 &  3.7 & -1.1 & 4.5 &   26.1 \\ 
   \hline
 & all & 2220 & -0.1 &  0.0 &  -6.6 &  3.6 & 1.1 &  1.0 & -0.9 & 6.3 & 1317.5 \\ 
   \hline
BMUS10Y & Regime 1 &  480 &  0.0 &  0.1 &  -2.5 &  2.0 & 1.0 &  0.6 & -0.3 & 3.2 &    5.7 \\ 
   & Regime 2 &  952 &  0.0 &  0.0 &  -3.7 &  3.0 & 1.2 &  0.8 &  0.0 & 3.3 &    3.2 \\ 
   & Regime 3 &  301 & -0.3 & -0.3 &  -4.4 &  4.8 & 2.3 &  2.7 &  0.2 & 2.9 &    2.4 \\ 
   & Regime 4 &  398 &  0.2 &  0.2 &  -3.1 &  3.1 & 1.4 &  1.0 & -0.2 & 3.2 &    3.6 \\ 
   & Regime 5 &   89 &  0.7 &  0.6 &  -5.6 &  6.7 & 3.5 &  6.2 &  0.1 & 2.8 &    0.4 \\ 
   \hline
 & all & 2220 &  0.0 &  0.0 &  -5.6 &  6.7 & 1.3 &  1.3 &  0.2 & 5.7 &  710.4 \\ 
   \hline
BMUS10Y GF & Regime 1 &  480 &  0.0 &  0.1 &  -3.0 &  2.1 & 1.1 &  0.7 & -0.2 & 3.2 &    5.4 \\ 
   & Regime 2 &  952 &  0.0 &  0.0 &  -4.2 &  3.5 & 1.2 &  0.8 & -0.1 & 3.7 &   22.5 \\ 
   & Regime 3 &  301 & -0.2 & -0.2 &  -3.7 &  3.0 & 1.6 &  1.5 &  0.0 & 2.9 &    0.1 \\ 
   & Regime 4 &  398 &  0.2 &  0.1 &  -3.0 &  3.9 & 1.3 &  0.9 &  0.0 & 3.5 &    4.6 \\ 
   & Regime 5 &   89 &  0.4 &  0.4 &  -3.4 &  4.1 & 2.2 &  2.6 &  0.0 & 2.7 &    0.4 \\ 
   \hline
 & all & 2220 &  0.0 &  0.0 &  -4.2 &  4.1 & 1.3 &  1.0 &  0.0 & 3.8 &   61.4 \\ 
   \hline
\end{tabular}
\endgroup
\end{table}
\Cref{fig:TS_5_states} shows the classification from the HMM model. 
\Cref{fig:LGC_five_States} shows the LGC maps for the 5 different regimes. The LGC is relatively uniform for regime 1 to regime 4 and the LGC coincide with the correlations/covariance that is identified in the HMM (see \Cref{Appendix_A}).
From the \Cref{tab:p_values_five_regiems} we observe that for the test between regime 1 and 4, the null hypothesis is accepted, i.e. the dependency is equal. Similarly we observe that regime 3 and regime 5 do not have a significantly different LGC structure. In this particular test setup, the number of pairwise tests is 10 and if we consider nominal level with $\alpha=0.05$, the Bonferroni correction would test each individual hypothesis at $\alpha^* = 0.05/10 = 0.005$. At a nominal level with $\alpha=0.01$, the null hypothesis between regime 2 and 3 also would be accepted. As can be viewed, the dependency test between regime 2 and 4 is barely within this threshold.   

With regards to LGC, this experiment has shown that in this particular case, five regime HMM seems to be an overparametrization with respect to the dependency structures and to the number of regimes. Although we observe more uniform LGC with more regimes, the dependency structures between several of the regimes are not significantly different. 

\begin{table}[!ht]
\caption{P-values five regimes.}\label{tab:p_values_five_regiems}
\centering
\begin{tabular}{llllll}
\toprule
\cmidrule(r){1-4}
P-value   & Regime 1  &  Regime 2 &  Regime 3 & Regime 4 & Regime 5  \\
\midrule
Regime 1      &  & 0 & 0 & \textbf{0.681} & 0 \\
Regime 2      &  &  & \textbf{0.004} & 0 & 0 \\
Regime 3      &  &  &  & 0 & \textbf{0.141} \\
Regime 4      &  &  &  &  & 0 \\
Regime 5      &  &  &  &  &  \\
\bottomrule
\end{tabular}
\end{table}


\section{Concluding remarks}

This paper presents a new procedure using the LGC for testing whether the dependency structures across different regimes, classified by a HMM, in financial returns are different. The test is a bootstrap procedure and the test statistic uses the squared difference of the estimated LGC between different regimes. With more than two regimes we have to perform a set of pairwise tests which in turn requires correction as the test expands and becomes a multi-comparison problem. 

The proposed test is verified by a study of significance level and power on simulated data. Both the level and power study are examined on 6 different models, all showing acceptable results. In addition, we conduct a simulation study where we classify observations with a HMM and compare how the misclassification affects the power of the test. By including the missclassified observations in the different regimes, the power decreases for all nominal levels. The power of the test is still on an acceptable level. The decrease in power is, however, an important aspect to have in mind during empirical analysis. 

We have illustrated the approach by performing empirical analysis on two different real data sets. First, we examine the daily returns on two stock indices, the S\&P500 and FTSE100, and second, stock market returns, S\&P500, against bond returns, the US 10 year government bond (US10Y). For the stock market indices, S\&P500 and FTSE100, we confirm that there are two regimes in the return series, a bull market regime with close to uniform dependency structure, and a bear market regime with overall higher dependence and an asymmetric structure, in particular, much higher dependence in the tails. Our approach thus confirms well known facts about the dependency structure between the S\&P500 and FTSE100 (see e.g \cite{Okimoto2008}). However, at the same time, we argue that our approach is more intuitive and easy to understand and interpret, than competing approaches. In particular, the LGC measure can be interpretable as the ordinary correlation. Furthermore, another main advantage is that in our framework we do not need a parametric assumption of the dependency structure in different regimes, and the proposed test can actually determine whether there are statistically significant differences between them.

For the stock-bond relationship, we considered weekly data and used the AIC and BIC to determine the optimal number of regimes in the HMM. The BIC and AIC showed that 3 and 5 regimes was preferred, respectively. We also assessed the classical two-regime model. The empirical analysis of returns of S\&P500 and US10Y showed that with increasing number of regimes, the LGC for the different regimes became more uniform and distinct. On the other hand we identified that the difference in dependency between several regimes was insignificant. Through our test, we showed and concluded that with respect to dependency, the five-regime HMM model was over-parameterized. In the three-regime model, the LGC documents a primarily positive relationship in the time period 1980-2000. From 2000 and onwards the relationship is mostly negative, whereas the regime associated with bear markets indicates less, but asymmetric dependence, hence documenting the loss of diversification benefits in times of crisis.

Although we have used Gaussian HMMs for simplicity, other more complex approaches for classification of observations is possible. In fact, because the analysis of dependency and the classification are separate procedures, we could use other more complex non-linear approaches such as neural networks or support vector machines or other state-of-the-art machine learning techniques for classification, see e.g. \citep{constantinou2006regime, hassan2007fusion, liu2019powering, mustafa2022detecting}. We leave this for future research. 

\paragraph*{Data availability statement}

The BMUS10Y is downloaded from Refinitin Eikon under the ticker \textit{.TRXVUSGOV10U}. This is the clean daily price index which has been aggregated to weekly observations for the purpose of this study. The other indices are publicly available and may be downloaded from e.g. Yahoo! Finance.

\paragraph*{Acknowledgements}
 This work was supported by the Financial Market Fund (Norwegian Research Council project no. 309218). We thank Dag Tjøstheim for valuable discussions and comments. 

\bibliography{references, hmms}
\bibliographystyle{abbrvnat}

\appendix

\section{Appendix}\label{Appendix_A}

\subsection{HMM Parameter Estimations}

\subsubsection{S\&P500 and FTSE100 Daily Returns 2 Regimes}

\pgfplotstableset{
every head row/.style={before row=\toprule,after row=\midrule},
every last row/.style={after row=\bottomrule}}


\begin{center}

\pgfplotstabletypeset[
every column/.code={
\ifnum\pgfplotstablecol=3
\pgfkeysalso{column type/.add={|}{}}%
\fi
\ifnum\pgfplotstablecol=6
\pgfkeysalso{column type/.add={|}{}}%
\fi
},
columns={Parameter, Estimate, Std.Error,Parameter, Estimate, Std.Error,Parameter, Estimate, Std.Error},
display columns/0/.style={select equal part entry of={0}{3},string type},
display columns/1/.style={select equal part entry of={0}{3},string type},
display columns/2/.style={select equal part entry of={0}{3},string type},
display columns/3/.style={select equal part entry of={1}{3},string type},
display columns/4/.style={select equal part entry of={1}{3},string type},
display columns/5/.style={select equal part entry of={1}{3},string type},
display columns/6/.style={select equal part entry of={2}{3},string type},
display columns/7/.style={select equal part entry of={2}{3},string type},
display columns/8/.style={select equal part entry of={2}{3},string type},
]{pgfplotstable.SP500_FTSE100_daily.dat}

\end{center}

\subsubsection{S\&P500 and US10Y Weekly Returns 2 Regimes}

\begin{center}
\pgfplotstabletypeset[
every column/.code={
\ifnum\pgfplotstablecol=3
\pgfkeysalso{column type/.add={|}{}}%
\fi
\ifnum\pgfplotstablecol=6
\pgfkeysalso{column type/.add={|}{}}%
\fi
},
columns={Parameter, Estimate, Std.Error,Parameter, Estimate, Std.Error,Parameter, Estimate, Std.Error},
display columns/0/.style={select equal part entry of={0}{3},string type},
display columns/1/.style={select equal part entry of={0}{3},string type},
display columns/2/.style={select equal part entry of={0}{3},string type},
display columns/3/.style={select equal part entry of={1}{3},string type},
display columns/4/.style={select equal part entry of={1}{3},string type},
display columns/5/.style={select equal part entry of={1}{3},string type},
display columns/6/.style={select equal part entry of={2}{3},string type},
display columns/7/.style={select equal part entry of={2}{3},string type},
display columns/8/.style={select equal part entry of={2}{3},string type},
]{pgfplotstable.SP500_US10Y_weekly_2_regimes.dat}
\end{center}

\subsubsection{S\&P500 and US10Y Weekly Returns 3 Regimes}

\begin{center}
\pgfplotstabletypeset[
every column/.code={
\ifnum\pgfplotstablecol=3
\pgfkeysalso{column type/.add={|}{}}%
\fi
\ifnum\pgfplotstablecol=6
\pgfkeysalso{column type/.add={|}{}}%
\fi
},
columns={Parameter, Estimate, Std.Error,Parameter, Estimate, Std.Error,Parameter, Estimate, Std.Error},
display columns/0/.style={select equal part entry of={0}{3},string type},
display columns/1/.style={select equal part entry of={0}{3},string type},
display columns/2/.style={select equal part entry of={0}{3},string type},
display columns/3/.style={select equal part entry of={1}{3},string type},
display columns/4/.style={select equal part entry of={1}{3},string type},
display columns/5/.style={select equal part entry of={1}{3},string type},
display columns/6/.style={select equal part entry of={2}{3},string type},
display columns/7/.style={select equal part entry of={2}{3},string type},
display columns/8/.style={select equal part entry of={2}{3},string type},
]{pgfplotstable.SP500_US10Y_weekly_3_regimes.dat}
\end{center}
\subsubsection{S\&P500 and US10Y Weekly Returns 5 Regimes}

\begin{center}
\pgfplotstabletypeset[
every column/.code={
\ifnum\pgfplotstablecol=3
\pgfkeysalso{column type/.add={|}{}}%
\fi
\ifnum\pgfplotstablecol=6
\pgfkeysalso{column type/.add={|}{}}%
\fi
},
columns={Parameter, Estimate, Std.Error,Parameter, Estimate, Std.Error,Parameter, Estimate, Std.Error},
display columns/0/.style={select equal part entry of={0}{3},string type},
display columns/1/.style={select equal part entry of={0}{3},string type},
display columns/2/.style={select equal part entry of={0}{3},string type},
display columns/3/.style={select equal part entry of={1}{3},string type},
display columns/4/.style={select equal part entry of={1}{3},string type},
display columns/5/.style={select equal part entry of={1}{3},string type},
display columns/6/.style={select equal part entry of={2}{3},string type},
display columns/7/.style={select equal part entry of={2}{3},string type},
display columns/8/.style={select equal part entry of={2}{3},string type},
]{pgfplotstable.SP500_US10Y_weekly_5_regimes.dat}
\end{center}

\subsection{GARCH Parameter Estimations}

For the GARCH(1,1) - filtrated data, that is S\&P500 daily, S\&P500 weekly, FTSE100 daily and BMUS10Y weekly data, we have used the following GARCH-model, with a Student $t$ error distribution, for the log-return $r_t$:
\begin{align*}
    r_t &= \mu + a_t, \\
    \alpha &= \sigma_t \epsilon_t, \\
    \sigma_t^2 &= \omega + \alpha a_{t-1}^2 + \beta \sigma_{t-1}^2
\end{align*}
where the notation is self explanatory. The standardised residuals are calculated as $\hat{a}_t = (r_t - \hat{\mu})/\hat{\sigma}_t$. Note that the conditional mean is just the mean of the observed log-returns $r_t$. The diagnostic output of the GARCH filtering is shown in the tables below, including the shape parameter of the Student $t$ error distribution.
\begin{table}[!ht]
    \centering
    \caption{S\&P500 Daily data}
    \begin{tabular}{rrrrr}
      \hline
     &  Estimate &  Std. Error &  t value & Pr($>|t|$) \\ 
      \hline
      $\mu$ & 0.070 & 0.007 & 9.602 & 0.000 \\ 
      $\omega$ & 0.011 & 0.002 & 5.403 & 0.000 \\ 
      $\alpha$ & 0.097 & 0.008 & 12.391 & 0.000 \\ 
      $\beta$ & 0.901 & 0.007 & 121.994 & 0.000 \\ 
      shape & 5.106 & 0.290 & 17.589 & 0.000 \\ 
       \hline
    \end{tabular}
    \label{tab:GARC_sp500_daily}
\end{table}

\begin{table}[!ht]
    \centering
    \caption{FTSE Daily data}
        \begin{tabular}{rrrrr}
        \hline
        &  Estimate &  Std. Error &  t value & Pr($>|t|$) \\ 
        \hline
        $\mu$ & 0.045 & 0.008 & 5.628 & 0.000 \\ 
        $\omega$ & 0.017 & 0.003 & 6.117 & 0.000 \\ 
        $\alpha$ & 0.094 & 0.008 & 11.671 & 0.000 \\ 
        $\beta$ & 0.891 & 0.009 & 99.135 & 0.000 \\ 
        shape & 7.518 & 0.550 & 13.668 & 0.000 \\ 
        \hline
        \end{tabular}
    
    \label{tab:GARCH_FTSE_daily}
\end{table}
\begin{table}[]
    \centering
    \caption{S\&P500 Weekly data} 
    \begin{tabular}{rrrrr}
      \hline
     &  Estimate &  Std. Error &  t value & Pr($>$$|$t$|$) \\ 
      \hline
      $\mu$ & 0.295 & 0.037 & 8.040 & 0.000 \\ 
      $\omega$ & 0.226 & 0.072 & 3.161 & 0.002 \\ 
      $\alpha$ & 0.142 & 0.027 & 5.303 & 0.000 \\ 
      $\beta$ & 0.821 & 0.034 & 24.374 & 0.000 \\ 
      shape & 5.634 & 0.650 & 8.664 & 0.000 \\ 
       \hline
    \end{tabular}
    \label{tab:GARCH_SP500_weekly_data}
\end{table}

\begin{table}[]
    \centering
    \caption{US10Y Weekly data} 
    \begin{tabular}{rrrrr}
      \hline
     &  Estimate &  Std. Error &  t value & Pr($>$$|$t$|$) \\ 
      \hline
      $\mu$ & 0.013 & 0.020 & 0.627 & 0.531 \\ 
      $\omega$ & 0.047 & 0.015 & 3.192 & 0.001 \\ 
      $\alpha$ & 0.095 & 0.017 & 5.692 & 0.000 \\ 
    $\beta$ & 0.868 & 0.024 & 35.865 & 0.000 \\ 
      shape & 10.000 & 1.835 & 5.450 & 0.000 \\ 
       \hline
    \end{tabular}
    \label{tab:GARCH_US10Y_weekly}
\end{table}

\newpage

\section{Appendix}\label{Appendix_B}

\subsection{Reparametrization of the likelihood
function}\label{reparametrization-of-the-likelihood-function}

In order to avoid dealing with constrained optimization, \(\bm{\psi}\) is parameterized into unconstrained ``working'' parameters, passed to an unconstrained optimizer, and the estimates are transformed back to retrieve original ``natural'' parameters. In particular, each natural covariance matrix is transformed into a working upper-triangular matrix using a Cholesky decomposition. Then, a log-transform of its diagonal elements allpws one to obtain an unconstrained parametrization, ensuring a positive definite symmetric natural covariance matrix. This is the ``log-Cholesky parametrization'' of \citet{pinheiro} and detailed in \citet[p.~260]{zucchini}. The means \(\mu_i\) are not modified, and the TPM elements \(\gamma_{ij}\) are transformed into \[ \tau_{ij} = \log\left(\frac{\gamma_{ij}}{1 - \sum_{k \neq i} \gamma_{ik}}\right) = \log(\gamma_{ij}/\gamma_{ii}), \text{ for } i \neq j. \] where \(\tau_{ij}\) are \(m(m-1)\) real-valued, thus unconstrained, elements of an \(m \times m\) matrix \(\textbf{T}\) with no diagonal elements. The diagonal elements of \(\bm{\Gamma}\) follows implicitly from \(\sum_j \gamma_{ij} = 1 \;\forall\; i\) \citep[p.~51]{zucchini}. The corresponding reverse transformation is given by \[ \gamma_{ij} = \frac{\exp(\tau_{ij})}{1 + \sum_{k \neq i} \exp(\tau_{ik})}, \text{ for } i \neq j, \] If necessary, the initial distribution \(\bm{\delta}\) can be reparametrized into \(\left(\log (\delta_2 / \delta_1), \log (\delta_3 / \delta_1), \ldots, \log (\delta_m / \delta_1) \right)\)

\subsection{Forward algorithm}\label{forward-algorithm}

The likelihood calculated in (\ref{eq:hmm_likelihood}) can be read as starting from the stationary distribution, then as a pass through the observations and changes in regime according to the Markov chain. This sets up an efficient evaluation method of the likelihood function: the so-called ``forward algorithm''. To set up this recursive algorithm, we define the vector \(\bm{\alpha}\) by
\begin{align*}
\bm{\alpha}_t &= \bm{\delta} \mathbf{P}(\bm{r_1})\bm{\Gamma} \mathbf{P}(\bm{r_2}) \bm{\Gamma} \mathbf{P}(\bm{r_3}) \ldots \bm{\Gamma} \mathbf{P}(\bm{r_t})\\
&= \bm{\delta} \mathbf{P}(\bm{r_1}) \prod_{s=2}^{t}\bm{\Gamma} \mathbf{P}(\bm{r_s})\\
&= \left( \alpha_t(1), \ldots, \alpha_t(m) \right)
\end{align*}
for \(t = 1, 2, \ldots, T\). The algorithm's name comes from its computation
\begin{gather*}
\bm{\alpha}_0 = \bm{\delta} \mathbf{P}(\bm{r_1})\\
\bm{\alpha}_t = \bm{\alpha}_{t-1} \bm{\Gamma} \mathbf{P}(\bm{r_t}) \text{ for } t = 1, 2, \ldots, T.
\end{gather*}
After passing through the observations, the likelihood is derived from
\begin{gather*}
L(\bm{\zeta)} = \bm{\alpha}_T \bm{1}'.
\end{gather*}

A scaled version of this algorithm (as described in \citet{zucchini}) is used to prevent numerical underflow errors.

\hypertarget{backward-algorithm}{%
\subsection{Backward algorithm}\label{backward-algorithm}}

Alternatively, the likelihood can be calculated with the help of the backward probabilities, defined similarly to the forward ones.
\begin{align*}
\bm{\beta}'_t &= \bm{\Gamma} \mathbf{P}(\bm{r_{t+1}}) \bm{\Gamma} \mathbf{P}(\bm{r_{t+2}}) \ldots \bm{\Gamma} \mathbf{P}(\bm{r_T}) \ldots \bm{1}'\\
&= \left(\prod_{s=t+1}^{T}\bm{\Gamma} \mathbf{P}(\bm{r_s}) \right) \bm{1}'\\
&= \left( \beta_t(1), \ldots, \beta_t(m) \right).
\end{align*}

The backward probabilities take their name from their recursive calculations \begin{gather*}
\bm{\beta}_T = \bm{1}'\\
\bm{\beta}_t = \bm{\Gamma} \mathbf{P}(\bm{r_{t+1}}) \bm{\beta}_{t+1} \text{ for } t = T-1, T-2, \ldots, 1.
\end{gather*}
The likelihood can also be calculated after passing through the observations \begin{gather*}
L(\bm{\zeta}) = \bm{\delta} \bm{\beta}_1.
\end{gather*}

\section{Appendix}\label{Appendix_C}

\begin{algorithm*}[ht]
	\begin{tabbing}
		\hspace{2em} \= \hspace{2em} \= \hspace{2em} \= \\
		{\bfseries Input}: \\
		\> - Returns: $\V R_t = \{R_{1t}, R_{2t} \}$ \\
		\> - Number of Regimes in HMM: $C$ \\
		\> - Discretization/grid: $\mathbf{x}_{ij}=(x_i, y_j)$ \\
		\> - Weight function: $w(x_i, y_j)$ \\
		\> - Bandwidths for LGC estimation: $\V b$ \\
		1 \> {\bfseries Fit a bivariate HMM on the two return series, } \\
		\> \> - Classify each observation into the most probable regime by finding the most probable regime \\
            \> \> \> $\argmax_{i \in \{1, \ldots, C \}} \text{P}(S_t = i \vert \V R_{(T)} = \V r_{(T)}) $ \\
        2 \> {\bfseries GARCH filtrate the returns} \\
		\> \> - Reduce the time dependence in the time series \\
		\> \> \> GARCH(1,1) $\leftarrow$ Fit a univariate GARCH model separately for both time series  \\
		3 \> {\bfseries Estimate the LGC map for both regimes} \\
		\> \>  - Specify $\mathbf{x}_{ij}$, $\V b$ and $w(x_i, y_j)$ \\
		\> \>  - Use GARCH filtrated returns from 2. \\
        \> \>  - For all $\mathbf{x}_{ij}$ in the grid, optimize the parameters of the LGC: \\
		\> \> \> $\bm \theta(\bm x) =\bm \theta(\mu_1(\bm x),\mu_2(\bm x),\sigma_1(\bm x),\sigma_2(\bm x),\rho(\bm x))$ \\
		4. \> {\bfseries Equality test of dependence} \\
		\> \>  - Bootstrap procedure with hypothesis \\
        \> \> \> $H_0: \quad \M \rho_{1}(x_i, y_j) = \M \rho_{2}(x_i, y_j) = \ldots = \M \rho_{C}(x_i, y_j) 
        \quad \text{for} \quad i,j=1,\cdots,n $\\
        \> \> \> $ H_1:\quad \M \rho_{1}(x_i, y_j) \neq \M \rho_{2}(x_i, y_j) \neq \ldots \neq \M \rho_{C}(x_i, y_j) \quad \text{for} \quad i,j=1,\cdots,n$ \\
        \> \> - with the following test statistic \\
		\> \> \> $ D_1^*(k,l) =
                \frac{1}{n^2}\sum\limits_{i=1}^{n} \sum\limits_{j=1}^{n} \left[ \hat{\M \rho}_{k}^*(x_i, y_j) -  \hat{\M \rho}_{l}^*(x_i, y_j) \right]^2  w(x_i, y_j) \quad \text{for} \quad k>l \quad \text{where} \quad k,l = 1 \ldots C $ \\
		\> \>  - If C > 2 $\rightarrow$ multiple comparison test problem  \\
		\> \> \> Adjust with e.g. Bonferroni correction. \\
		{\bfseries return}  p-value(s)  	
	\end{tabbing}	
	\vspace{0.001cm}
	\caption{Algorithm to test for asymmetric dependence across regimes.}
	\label{BCNN_decision_alg}
\end{algorithm*}

\end{document}